\documentclass[%
aip,
  amsmath,amssymb,
  reprint,%
  floatfix,
]{revtex4-1}

\usepackage{graphicx}
\usepackage{dcolumn}
\usepackage{bm}
\usepackage{subfigure}
\usepackage{xcolor}
\usepackage{comment}
\usepackage[normalem]{ulem}

\usepackage[utf8]{inputenc}
\usepackage[T1]{fontenc}
\usepackage{mathptmx}
\usepackage[colorlinks=true, allcolors=blue]{hyperref}
\usepackage{url}
\usepackage{lineno}
\usepackage{wasysym} 

\renewcommand{\thefootnote}{\roman{footnote}}
\newcommand\orcid[1]{\href{https://orcid.org/#1}{\includegraphics[height=9pt]{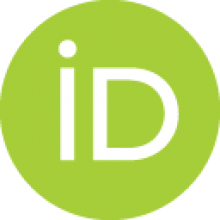}}}

\newcommand{\nocontentsline}[3]{}
\newcommand{\tocless}[2]{\bgroup\let\addcontentsline=\nocontentsline#1{#2}\egroup}

\begin{document}

\preprint{AIP/123-QED}
\title{The Simons Observatory: Cryogenic Half Wave Plate Rotation Mechanism for the Small Aperture Telescopes}

\author{K.~Yamada~\orcid{0000-0003-0221-2130}}
\email[Authors to whom correspondence should be addressed:
\href{mailto: ykyohei@cmb.phys.s.u-tokyo.ac.jp}{ykyohei@cmb.phys.s.u-tokyo.ac.jp}, 
\href{yuki.sakurai727@gmail.com}{yuki.sakurai727@gmail.com}, 
\href{ysakurai@s.okayama-u.ac.jp}{ysakurai@s.okayama-u.ac.jp}]{ }
\affiliation{Department of Physics, Graduate School of Science, The University of Tokyo, Tokyo 113-0033, Japan\looseness=-1}
\author{B.~Bixler}
\affiliation{Department of Physics, University of California, San Diego, La Jolla, CA 92093, USA}
\author{Y.~Sakurai~\orcid{0000-0001-6389-0117}}
\email[Authors to whom correspondence should be addressed:
\href{mailto: ykyohei@cmb.phys.s.u-tokyo.ac.jp}{ykyohei@cmb.phys.s.u-tokyo.ac.jp}, 
\href{yuki.sakurai727@gmail.com}{yuki.sakurai727@gmail.com}, 
\href{ysakurai@s.okayama-u.ac.jp}{ysakurai@s.okayama-u.ac.jp}]{ }

\affiliation{Graduate school of natural science and technology, Okayama University, Okayama 700-8530 Japan}
\affiliation{Kavli Institute for the Physics and Mathematics of the Universe (WPI), UTIAS, The University of Tokyo, Chiba 277-8583, Japan\looseness=-1}
\author{P.~C.~Ashton}
\thanks{Current address: Advanced Technology and Systems Division, SRI International, Menlo Park, CA 94025, USA}
\affiliation{Physics Division, Lawrence Berkeley National Laboratory, Berkeley, CA 94720, USA}
\affiliation{Department of Physics, University of California, Berkeley, CA 94720, USA}
\affiliation{Kavli Institute for the Physics and Mathematics of the Universe (WPI), UTIAS, The University of Tokyo, Chiba 277-8583, Japan\looseness=-1}

\author{J.~Sugiyama~\orcid{0009-0007-7435-9082}}
\affiliation{Department of Physics, Graduate School of Science, The University of Tokyo, Tokyo 113-0033, Japan\looseness=-1}
\author{K.~Arnold~\orcid{0000-0002-3407-5305}}
\affiliation{Department of Physics, University of California, San Diego, La Jolla, CA 92093, USA}
\author{J.~Begin~\orcid{0000-0003-2607-4676}}
\affiliation{Joseph Henry Laboratories of Physics, Jadwin Hall, Princeton University, Princeton, NJ 08544, USA\looseness=-1}
\author{L.~Corbett}
\affiliation{Department of Physics, University of California, Berkeley, CA 94720, USA}
\affiliation{Physics Division, Lawrence Berkeley National Laboratory, Berkeley, CA 94720, USA}
\author{S.~Day-Weiss~\orcid{0009-0003-5814-2087}}
\affiliation{Joseph Henry Laboratories of Physics, Jadwin Hall, Princeton University, Princeton, NJ 08544, USA\looseness=-1}
\author{N.~Galitzki~\orcid{0000-0001-7225-6679}}
\affiliation{Department of Physics, University of Texas at Austin, Austin, TX 78722, USA}
\affiliation{Weinberg Institute for Theoretical Physics, Texas Center for Cosmology and Astroparticle Physics, Austin, TX 78712, USA\looseness=-1}
\author{C.~A.~Hill~\orcid{0000-0002-2641-6878}}
\affiliation{Department of Physics, University of California, Berkeley, CA 94720, USA}
\affiliation{Physics Division, Lawrence Berkeley National Laboratory, Berkeley, CA 94720, USA}
\author{B.~R.~Johnson~\orcid{0000-0002-6898-8938}}
\affiliation{Department of Astronomy, University of Virginia, Charlottesville, VA 22904, USA}
\author{B.~Jost~\orcid{0000-0002-0819-751X}}
\affiliation{Kavli Institute for the Physics and Mathematics of the Universe (WPI), UTIAS, The University of Tokyo, Chiba 277-8583, Japan\looseness=-1}
\author{A.~Kusaka~\orcid{0009-0004-9631-2451}}
\affiliation{Physics Division, Lawrence Berkeley National Laboratory, Berkeley, CA 94720, USA}
\affiliation{Department of Physics, Graduate School of Science, The University of Tokyo, Tokyo 113-0033, Japan\looseness=-1}
\affiliation{Kavli Institute for the Physics and Mathematics of the Universe (WPI), UTIAS, The University of Tokyo, Chiba 277-8583, Japan\looseness=-1}
\affiliation{Research Center for the Early Universe, School of Science, The University of Tokyo, Tokyo 113-0033, Japan\looseness=-1}
\author{B.~J.~Koopman~\orcid{0000-0003-0744-2808}}
\affiliation{Wright Laboratory, Department of Physics, Yale University, New Haven, CT 06520, USA\looseness=-1}
\author{J.~Lashner~\orcid{0000-0002-6522-6284}}
\affiliation{Wright Laboratory, Department of Physics, Yale University, New Haven, CT 06520, USA\looseness=-1}
\author{A.~T.~Lee~\orcid{0000-0003-3106-3218}}
\affiliation{Department of Physics, University of California, Berkeley, CA 94720, USA}
\affiliation{Physics Division, Lawrence Berkeley National Laboratory, Berkeley, CA 94720, USA}
\author{A.~Mangu}
\affiliation{Department of Physics, University of California, Berkeley, CA 94720, USA}
\author{H.~Nishino~\orcid{0000-0003-0738-3369}}
\affiliation{Research Center for the Early Universe, School of Science, The University of Tokyo, Tokyo 113-0033, Japan\looseness=-1}
\author{L.~A.~Page~\orcid{0000-0002-9828-3525}}
\affiliation{Joseph Henry Laboratories of Physics, Jadwin Hall, Princeton University, Princeton, NJ 08544, USA\looseness=-1}
\author{M.~J.~Randall}
\affiliation{Department of Physics, University of California, San Diego, La Jolla, CA 92093, USA}
\author{D.~Sasaki~\orcid{0009-0003-2513-2608}}
\affiliation{Department of Physics, Graduate School of Science, The University of Tokyo, Tokyo 113-0033, Japan\looseness=-1}
\author{X.~Song}
\affiliation{Department of Physics, University of California, Berkeley, CA 94720, USA}
\author{J.~Spisak~\orcid{0000-0003-1789-8550}}
\affiliation{Department of Physics, University of California, San Diego, La Jolla, CA 92093, USA}
\author{T.~Tsan~\orcid{0000-0002-1667-2544}}
\affiliation{Department of Physics, University of California, San Diego, La Jolla, CA 92093, USA}
\author{Y.~Wang~\orcid{0000-0002-8710-0914}}
\affiliation{Joseph Henry Laboratories of Physics, Jadwin Hall, Princeton University, Princeton, NJ 08544, USA\looseness=-1}
\author{P.~A.~Williams~\orcid{0000-0003-3920-7669}}
\affiliation{Physics Division, Lawrence Berkeley National Laboratory, Berkeley, CA 94720, USA}

\date{\today}

\begin{abstract}
We present the requirements, design and evaluation of the cryogenic continuously rotating half-wave plate (CHWP) for the Simons Observatory (SO).
SO is a cosmic microwave background (CMB) polarization experiment at Parque Astron\'{o}mico Atacama in northern Chile that covers a wide range of angular scales using both small ($\diameter$0.42~m) and large ($\diameter$6~m) aperture telescopes.
In particular, the small aperture telescopes (SATs) focus on large angular scales for primordial B-mode polarization.
To this end, the SATs employ a CHWP to modulate the polarization of the incident light at 8~Hz, suppressing atmospheric $1/f$ noise and mitigating systematic uncertainties that would otherwise arise due to the differential response of detectors sensitive to orthogonal polarizations.
The CHWP consists of a 505~mm diameter achromatic sapphire HWP and a cryogenic rotation mechanism, both of which are cooled down to $\sim$50~K to reduce detector thermal loading.
Under normal operation the HWP is suspended by a superconducting magnetic bearing and rotates with a constant 2~Hz frequency, controlled by an electromagnetic synchronous motor.  
We find that the number of superconductors and magnets that make up the superconducting magnetic bearing are important design parameters, especially for the rotation mechanism's vibration performance. 
The rotation angle is detected through an angular encoder with a noise level of 0.07\,$\mu$rad$\sqrt{\mathrm{s}}$.
During a cooldown, the rotor is held in place by a grip-and-release mechanism that serves as both an alignment device and a thermal path.
In this paper we provide an overview of the SO SAT CHWP: its requirements, hardware design, and laboratory performance.

\end{abstract}

\maketitle

\tableofcontents
\section{Introduction}
\label{sec:into}

The cosmic microwave background radiation (CMB) is the oldest detectable light in the universe, originating from the epoch of recombination. 
Its polarization is dominated by parity-even "E-mode" patterns, primarily sourced by density fluctuations in the early universe. \cite{kamionkowski_statistics_1997, zaldarriaga_all-sky_1997}
Primordial tensor perturbations are predicted to 
produce parity-odd "B-mode" polarization with an angular spectrum peaking at degree scales.\cite{kamionkowski_probe_1997, seljak_signature_1997} 
The rapid expansion of the early universe, called inflation, could produce tensor perturbations. \cite{abbott1986inflationary, linde2005particle, Linde2007}

A measurement of the primordial B-mode polarization signature would constrain models of the early universe and contribute to the understanding of physics at grand unified theory (GUT) energy scales.\cite{seljak_direct_1999}
The primordial B-mode signal is subdominant to both E-modes and several foreground sources, including
polarized galactic emission, mainly from synchrotron and thermal dust,\cite{Page_2007, planck_collaboration_planck_2018} and E to B-mode conversion through gravitational lensing. \cite{zaldarriaga_gravitational_1998}
Separating primordial B-mode polarization from galactic foregrounds requires wide frequency coverage, while efficient lensing B-mode separation requires high resolution and large sky coverage.\cite{knox_limit_2002, kesden_separation_2002, seljak_gravitational_2004}

The Simons Observatory (SO) is a CMB experiment located at Cerro Toco (5,200~m) in the Chilean Atacama Desert. \cite{Ade2019, arxiv.1907.08284}
The nominal observatory consists of three small aperture telescopes (SATs), each with an aperture of 42~cm targeting large angular scales,\cite{Ali2020, Kiuchi2020} and one large aperture telescope (LAT), with an aperture of 6~m targeting arcminute angular scales.\cite{galitzki_simons_2018, Zhu_2021, LAT_CCAT_2018}
The combined arrays of the SATs and the LAT employ over 60,000 transition-edge sensor (TES) bolometers. \cite{10.1117/12.2314435,li2020,McCarrick_2021}
The SAT are designed for the primary science goal of constraining primordial B-modes and therefore probe a sufficiently wide range of frequency bands and angular scales.
In order to address these requirements, each SAT is sensitive to large angular scales of 30 $<$ $\ell$ $<$ 300 with 10\% fractional sky coverage and has two primary frequency bands with bandwidths ranging from about 20\% to 45\% of the central frequencies.\cite{bolocalc-so-model}
Two SATs cover middle frequencies (MF), with band centers at of 93 and 145~GHz, while the third SAT covers ultra-high frequencies (UHF), centering at 225 and 280~GHz.

Various modulation techniques have been used in CMB polarization experiments. \cite{Map_radiometer_2003, POLAR_2003, DASI_2005, Barkats_2005, Chen_2009, planck_lf_instrument, QUIET_2012, Moyerman_2013, Bryan_2016, Miller_2016, GB2020, Harrington_2021} The rapid modulation of incident polarization by a half-wave plate (HWP) is one of the most promising techniques to separate the large scale CMB B-mode polarization signal from large unpolarized atmospheric $1/f$ noise, while using an instrument with large optical throughput and a large number of multi-chroic pixels.\cite{johnson_maxipol_2007, matsumura_phd, klein_cryogenic_2011, kusaka_modulation_2014, hill_design_2016, johnson_large-diameter_2017, HillPB2bCHWP2020}

The SO SATs employ a cryogenic continuously rotating half-wave plate (CHWP) polarization modulator to achieve sensitive observations at large angular scales.
In this paper, we describe the requirements, design, and evaluation test results for the SO SAT CHWP rotation mechanism for the MF and UHF frequency bands. The development and study for low frequencies (LF), having center band frequencies of 27 and 39~GHz, is underway and will be reported elsewhere.
Section \ref{sec:principle} provides an overview of the basic functionality of the polarization modulator,
Sec. \ref{sec:hwp_requirements} presents the requirements, 
Sec. \ref{sec:chwp_design} presents the design of subsystems,
Sec. \ref{sec:performance} presents the results of laboratory performance, 
and Sec. \ref{sec:conclusion} summarizes the development and describes future prospects.


\begin{figure*}[t]
 \begin{center}
    \includegraphics[width = 0.95\textwidth]{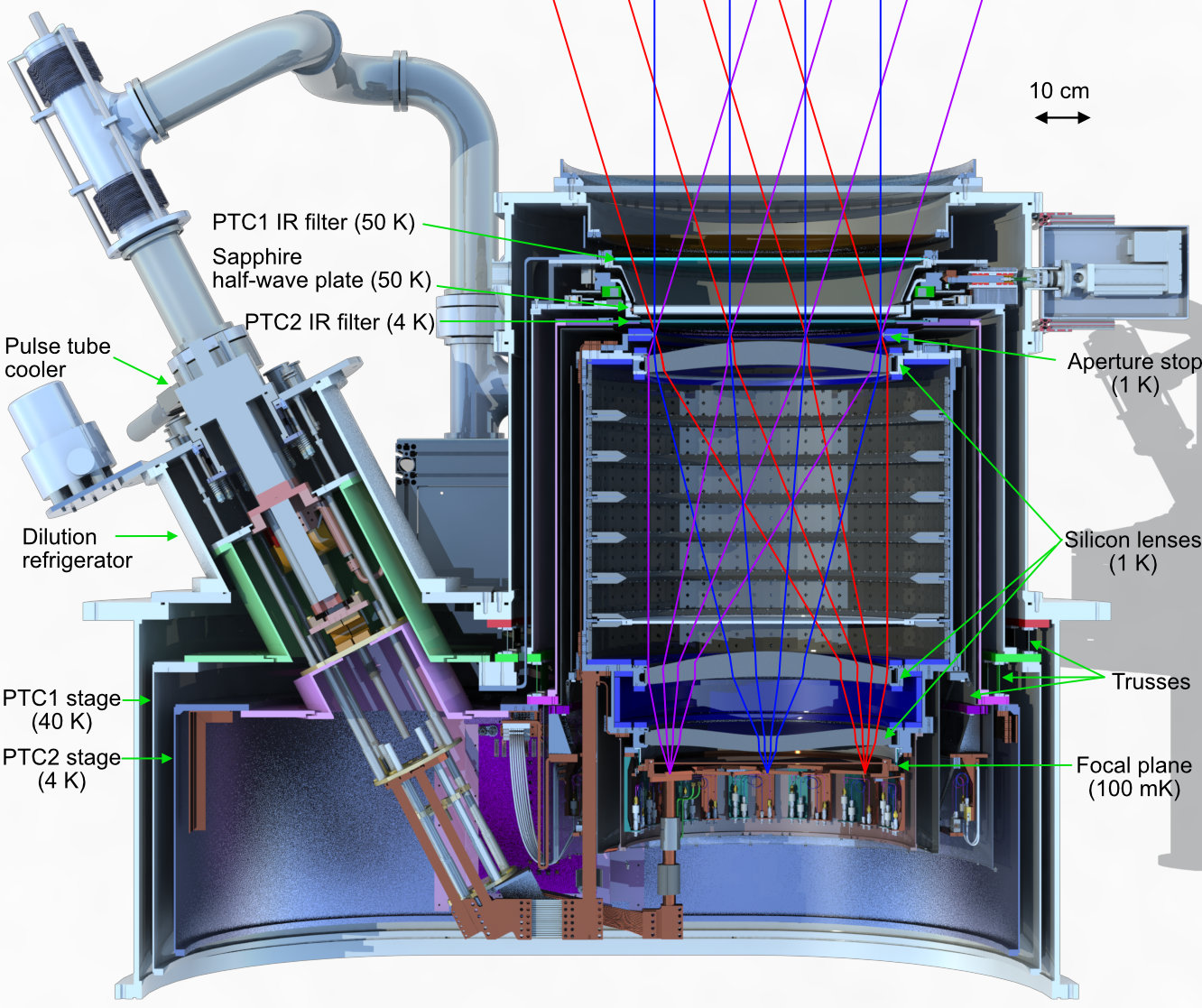}
    \includegraphics[width = 0.95\textwidth]{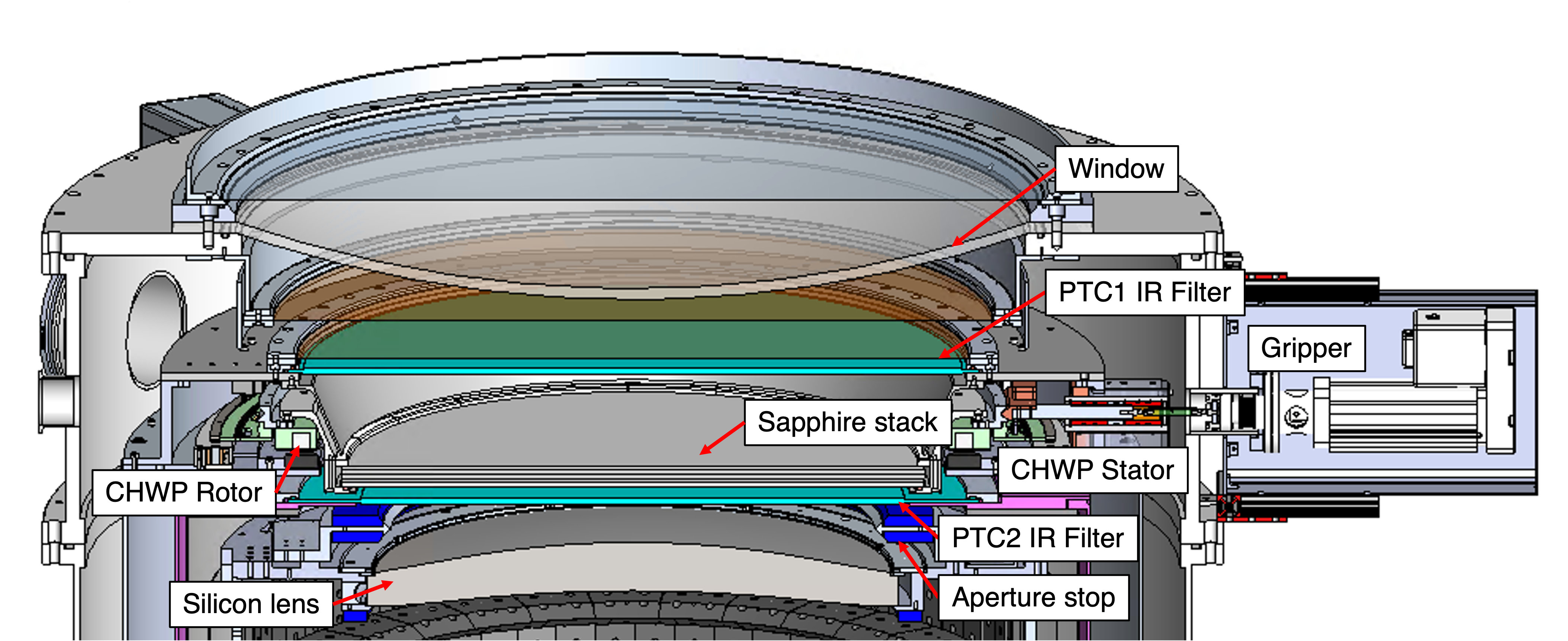}
    \caption{The computer-aided design (CAD) cross-section of the small-aperture telescope's receiver.\cite{galitzki_simons_2018} The superposed optical ray traces\cite{10.1117/12.2561205} are approximations and for illustrative purposes only. The SAT receiver employs an outer vacuum shell and two thermal shells cooled by two pulse tube coolers (PTCs). We refer to the two thermal shells as the PTC1 stage and the PTC2 stage. 
    The CHWP rotation mechanism and PTC1 infrared (IR) blocking alumina filter are located on the PTC1 stage, and the PTC2 IR blocking alumina filter is located on the PTC2 stage. The temperatures shown for each component are nominal values.}
    \label{fig:design_cad_sat}
 \end{center}
\end{figure*}

\section{Half-wave plate polarimetry}
\label{sec:principle}
Observing primordial $B$-modes requires 
nK sensitivity, especially at degree angular scales. 
The challenge for ground-based polarimeters is to characterize faint signals in the presence of comparably large-amplitude unpolarized atmospheric fluctuations, in addition to polarized emission from the atmosphere,\cite{Takakura2019, Petroff2020} the ground, and the instrument itself.
While atmospheric noise is mitigated by the high altitude and low water vapor of the SO site, it is nevertheless difficult to control this contamination due to spatial and temporal fluctuations caused by local weather conditions. The atmospheric noise enters the detector time ordered data (TOD) as $1/f$ noise, reducing sensitivity at large angular scales. \cite{church_atm, Dunner_2013, Errard2015, PhysRevD.105.042004}

Experiments without rapid polarization usually mitigate the atmospheric $1/f$ noise and reconstruct the linear polarization by differencing detectors with orthogonal antennas. \cite{BICEP_characterization_2010, ACT_overview_2011, SPTpol2012}
However, any mismatched response between orthogonal detectors leads to intensity to polarization (I-to-P) leakage, which contaminates the cosmological polarization signal. \cite{shimon_cmb_2008}

In order to mitigate both $1/f$ noise and I-to-P leakage, the SATs employ a CHWP-based polarization modulation system, a commonly used technique among millimeter and sub-millimeter polarimeters. The HWP consists of a birefringent material that introduces a phase difference of 180$^\circ$ between the ordinary and extraordinary axes. 
While passing through the HWP, the incident polarization rotates by twice the angle between its vector and the HWP extraordinary axis. 
If the HWP is continuously rotating, the measured linear polarization signal rotates synchronously, and is modulated above $1/f$ noise fluctuations by setting the appropriate rotation frequency.
The HWP modulated signal incident on a polarimeter sensitive to a single linear polarization is expressed as
\begin{equation}
    d_{m}(t) = I(t) + \epsilon \mathrm{Re} [\left(Q(t) + iU(t)\right)m(\chi)],
\end{equation}
where $t$ is time, $I$, $Q$, and $U$ are the Stokes vectors of the incident light, assuming $V=0$. $\epsilon$ is the HWP polarization modulation efficiency, $\chi(t)$ is the rotation angle of the HWP, and 
\begin{equation}
    m(\chi) = \exp(-i4\chi)
\end{equation}
is the modulation function. 
To extract the intensity signal, we low-pass filter the modulated data.
To extract the linear polarization signals, $Q$ and $U$, we band-pass filter the modulated data around the modulation frequency, multiply by the complex conjugate of the modulation function, $m(\chi)$, and apply a low-pass filter to it.
We call this procedure demodulation. \cite{kusaka_modulation_2014} 
There is no need for differencing pairs of orthogonal detectors in demodulation, and thus all functional TESs in the array can operate independently.

The CHWP employs a Pancharatnam-style\citep{pancharatnam_achromatic_1955} achromatic HWP (AHWP) composed of three single-crystal sapphire plates. Sapphire was chosen because it has high thermal conductivity and low millimeter-wave absorption. The AHWP allows coverage of a wide frequency band, and it has been employed in several previous CMB experiments\cite{the_ebex_collaboration_ebex_2018, johnson_maxipol_2007, ModelingHWPBlastpol2013, Bryan_2016, HillPB2bCHWP2020} as well as been included in the design of planned instruments. \cite{LB_PMU_LF}
The design of the AHWP is similar to that used in the Simons Array experiment, including the 3 slabs of sapphire with anti-reflection (AR) coating. \cite{hill_design_2016}
The control of systematic effects from the AHWP is a crucial aspect of HWP polarimetry.\cite{ModelingHWPBlastpol2013, essinger-hileman_systematic_2016, Monelli_2023}
The study of detailed performance and systematics of the SO AHWP is reported in \textcite{SO_AHWP}  
The design of the rotation mechanism inherits many aspects of the analogous system in the Simons Array experiment.\cite{HillPB2bCHWP2020} 
However, the details of the implementation have been modified according to the unique requirements of the SO SAT receiver, as shown in Fig. \ref{fig:design_cad_sat}. 

The driving philosophy behind the CHWP system design is to ensure stability, both mechanically and thermally. Since the CHWP is located in the SAT's main beam (Fig. \ref{fig:design_cad_sat}), any unpredictable behavior carries a risk of contaminating the polarization signal from the sky.\cite{10.1117/12.2561205}
Therefore, we require any variation in the CHWP system other than its fast modulation to be small, slowly varying, and measurable where possible. Furthermore, we look for robust design solutions that will withstand years of continuous operation at the Simons Observatory site while minimizing maintenance and telescope downtime. 
Finally, we develop the CHWP rotation mechanism as a modular system that can be tested with more frequent iterations before being integrated with the complete SAT cryostat. 
The SAT receiver employs an outer vacuum shell and two thermal shells cooled by two pulse tube coolers (PTCs).\cite{PTC} 
We refer to the two thermal shells as the PTC1 stage and the PTC2 stage with nominal temperatures of 40~K and 4~K, respectively. The CHWP mechanism is mounted on the most skyward flange of the PTC1 stage (Fig.~\ref{fig:design_cad_sat}).


\section{Requirements}
\label{sec:hwp_requirements}
\begin{table}
\caption{The CHWP's numerical requirements and achieved values.}
\label{tab:requirements}
\begin{ruledtabular}
\begin{tabular}{l l l}
\textbf{Parameter} & \textbf{Requirement} & \textbf{Achieved} \\
\hline
Assembly outer diameter & $\leq$ 950~mm & 931~mm \\
\hline
Assembly height & $\leq$ 246~mm & 124.5~mm \\
\hline
Cryogenic stage mass & $\leq$ 70~kg & 66~kg \\
\hline
Clear aperture diameter & $\geq$ 478~mm & 490~mm \\
\hline
Rotor center alignment & $\leq 5$~mm & $\leq 4.5$~mm \\
\hline
Rotor temperature $T_{\mathrm{rotor}}$ & $\leq$ 85~K & $\leq$ 70~K \\
\hline
Stator temperature $T_{\mathrm{stator}}$ & $\leq$ 70~K & $\leq$ 60~K\footnote{Achieved at steady state when the rotor rotates at 2~Hz (Fig.~\ref{fig:thermal_profile}).} \\
\hline
Thermal dissipation $P_{\mathrm{stator}}$ & $\leq$ 3~W & $\leq$ 1.6~W\\ 
\hline
Rotor thermalization time\footnote{Lag time of rotor vs stator on initial cooldown (Fig.~\ref{fig:rotor_thermalization})} & $\leq$ 36~hr & 10~hr \\
\hline
Rotation frequency $f_{\mathrm{HWP}}$ & 2~Hz & 0.5~Hz - 3~Hz \\
\hline
Rotation stability $\Delta f_{\mathrm{HWP}}$ \footnote{The requirement is over the observation period of several years. The achieved stability is over 4 days (Fig. \ref{fig:stability}).} & $\leq 10~\mathrm{mHz}$ & $\leq 5~\mathrm{mHz}$\\
\hline
Encoded angle noise & $\ll$ 3 $\mu\mathrm{rad}\sqrt\mathrm{s}$ & 0.07~$\mu\mathrm{rad}\sqrt\mathrm{s}$\\
\end{tabular}
\end{ruledtabular}
\end{table}

Table~\ref{tab:requirements} summarizes the CHWP requirements and achieved values.
The system design relies on the heritage of the CHWP for the Simons Array telescopes,\cite{HillPB2bCHWP2020} and the requirements qualitatively remain the same. 
Notable differences to the requirements include: a) a larger optical throughput requiring optical diameter of 490~mm and the placement of the HWP optics close to the telescope's aperture stop; b) relaxed requirements on the physical volume available to the system; and c) a larger variation of the gravity vector due to the addition of bore sight rotation about the SAT’s optical axis to the scan strategy
resulting in the rotor alignment requirement.

\subsection{Operational}
During normal operation we require a steady rotation frequency ($f_\mathrm{HWP}$) of 2 Hz or greater in order for the polarization signal to be modulated faster than the $1/f$ knee of temperature fluctuations in the atmosphere. \cite{kusaka_modulation_2014, takakura_performance_2017}
In order to make efficient use of observing time, we require spin-up (down) to (from) this rotation frequency to be achieved in less than 5~minutes. 
We also require the peak-to-peak stability of $f_\mathrm{HWP}$ to be better than 10~mHz during observations over a standard observation period of several years. 
Requirements are determined by operational conditions such as the need to avoid noisy frequency regions in the TES power spectra or resonance frequencies of the mechanical structures in the cryostat.

\subsection{Mechanical}
\label{req-mech}
For modularity, the CHWP rotation mechanism is designed to easily accommodate the front end, window-side, of the cryostat. This constrains the total envelope of the CHWP cryogenic components to be $<950$~mm in diameter and $<246$~mm in height along the optical axis.
The total mass requirement of the CHWP, $\leq 70$~kg, is not tightly constrained, as it is subdominant to the total suspended mass of the cryogenic stages supported by the primary cryo-mechanical truss. \cite{SOtruss}


CHWP rotation can generate vibrations in the cryostat, and heat up the focal plane’s temperature, degrading detector gain stability and introducing $1/f$ noise. 
We require no measurable CHWP-induced resonant heating in the focal plane temperature.
The vibrational performance is discussed in Sec.~\ref{perf-vib}.

\subsection{Optical} \label{req-opt}

\begin{figure}
    \centering
    \includegraphics[width = 0.45\textwidth]{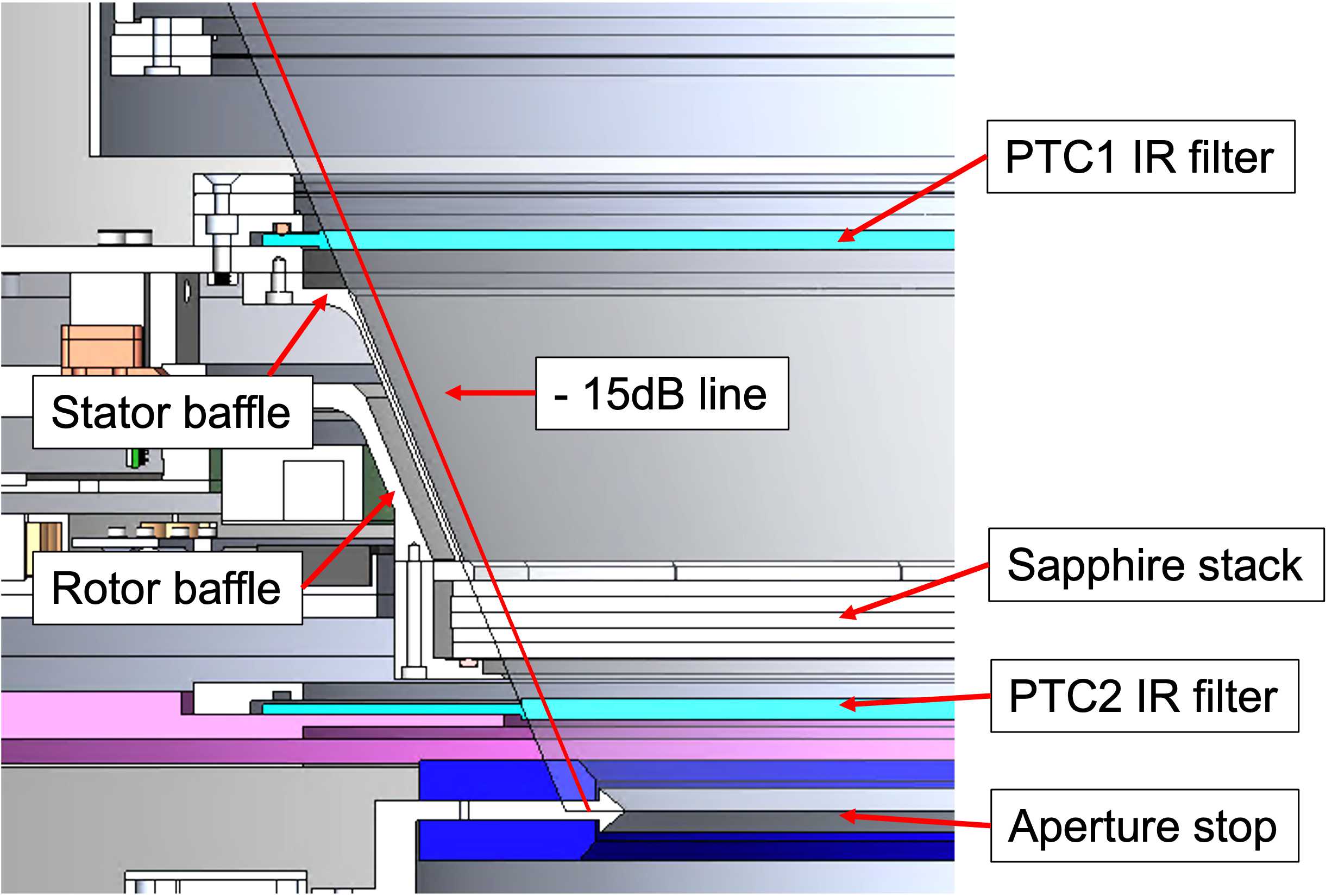}
    \caption{A cross-sectional view of the CHWP system with the approximation of the $-15$~dB line of the 90~GHz beam, determined based on physical-optics simulations.\cite{10.1117/12.2561205}}
    \label{fig:cradle}
\end{figure}

A clear aperture is one of the central requirements for the CHWP system. Special care is taken to prevent any optical interference from moving parts of the rotation mechanism, as any signal modulated at $4 f_\mathrm{HWP}$ mimics incident polarization. 
To this end, we require all parts other than the sapphire stack to be circularly symmetric, and to be outside of the $-15$~dB line of the 90~GHz beam (Fig.~\ref{fig:cradle}). 
The total power of the beam below the $-15$~dB level is estimated to be 0.52\%. 
The $-15$~dB requirement is determined by a physical-optics simulation to ensure that sidelobe scattering from the CHWP components is small compared to that from other optical elements of SAT.\cite{10.1117/12.2561205} 
The 90\,GHz band is chosen to define the $-15$~dB line because it gives the most stringent optical requirements compared to the other MF and UHF bands. 

Given the SAT's $35^\circ$ field of view and the diffraction limited beam, these requirements impose an approximated keepout zone with an opening angle of about 23$^{\circ}$ extending skyward from the 420-mm diameter aperture stop (Fig.~\ref{fig:cradle}). 
The sapphire stack must be at least 490~mm in diameter at a distance of 40~mm away from the aperture stop; as close to the stop as possible while maintaining sufficient mechanical margin for the PTC2 alumina filter in between.

Finally, the misalignment of the rotor with the optical center must not exceed 5~mm during operation,
as any larger misalignment would lead to physical interference or contamination of the beam.

\subsection{Thermal}
\label{req-therm}
The rotor temperature is required to stay below 85~K to reduce thermal emission, requiring the thermal loading by the CHWP induced IR radiation incident on the PTC2 alumina filter to be subdominant to the total thermal loading on the PTC2 stage.\cite{SAT-MF1} 
A lower temperature of the HWP optics reduces the fluctuation and non-uniformity of its emissivity and also helps to reduce the potential systematics in the observed polarization signal.

The stator temperature is required to be below 70~K, which is well below the critical temperature of the superconducting bearing, 95~K (Secs.~\ref{des-smb} and \ref{perf-sag}).
The power dissipation on the PTC1 stage is required to be subdominant to the total thermal loading on the PTC1 stage, and also care is taken to effectively heat sink the CHWP stator to the PTC1 stage to reduce temperature gradients generated by the loading from the CHWP system during operation, and to improve drive system efficiency by reducing the resistance of the motor drive coils. As such, we keep the thermal load on the PTC1 stage below $\sim 3$~W during operation.
The low-power dissipation is an advantage not only for the cryogenic components, but also for the room-temperature components, especially for the experiments conducted at the altitude of 5,200 m, where convective air cooling is poor.


\subsection{Data Acquisition} \label{req-daq}
Precise measurement of the rotation angle of the HWP is crucial for demodulation in the analysis pipeline. 
We require that the propagated noise equivalent temperature (NET) of the angular encoder be an order of magnitude smaller than the MF SAT NET goal: 1.4~$\mu\mathrm{K}\sqrt\mathrm{s}$,\cite{Ade2019} where we combine the sensitivity at 93 and 145~GHz to be conservative.  
Assuming a $\sim 100~\mathrm{mK}$ constant polarization induced by the vacuum window and the PTC1 alumina filter,\cite{salatino_studies_2018} and following a similar calculation to that described in Hill \textit{et~al.},\cite{HillPB2bCHWP2020} this imposes a limit of 3 $\mu\mathrm{rad}\sqrt\mathrm{s}$ for the encoder white noise level.
In addition to the low encoding noise, robustness in encoding, such as a low rate of data packet loss and of glitches, is crucial. \cite{PBlowell_2022}

Due to the remote nature of the SO site, we also require that the CHWP system be capable of autonomous operation, including the capability of its diagnostic monitoring system to detect a power interruption to the cryostat and trigger an automatic shutdown procedure, braking and re-gripping of the rotor.

\section{Design}
\label{sec:chwp_design}
This section describes the mechanical design of the rotation mechanism and highlights its novel aspects. The overall mechanism can be divided into five major components: the HWP, the superconducting magnetic bearing (SMB), the grippers, the motor, and the angle encoder. 

\begin{figure*}
    \begin{center}
    \includegraphics[width = 0.95\textwidth]{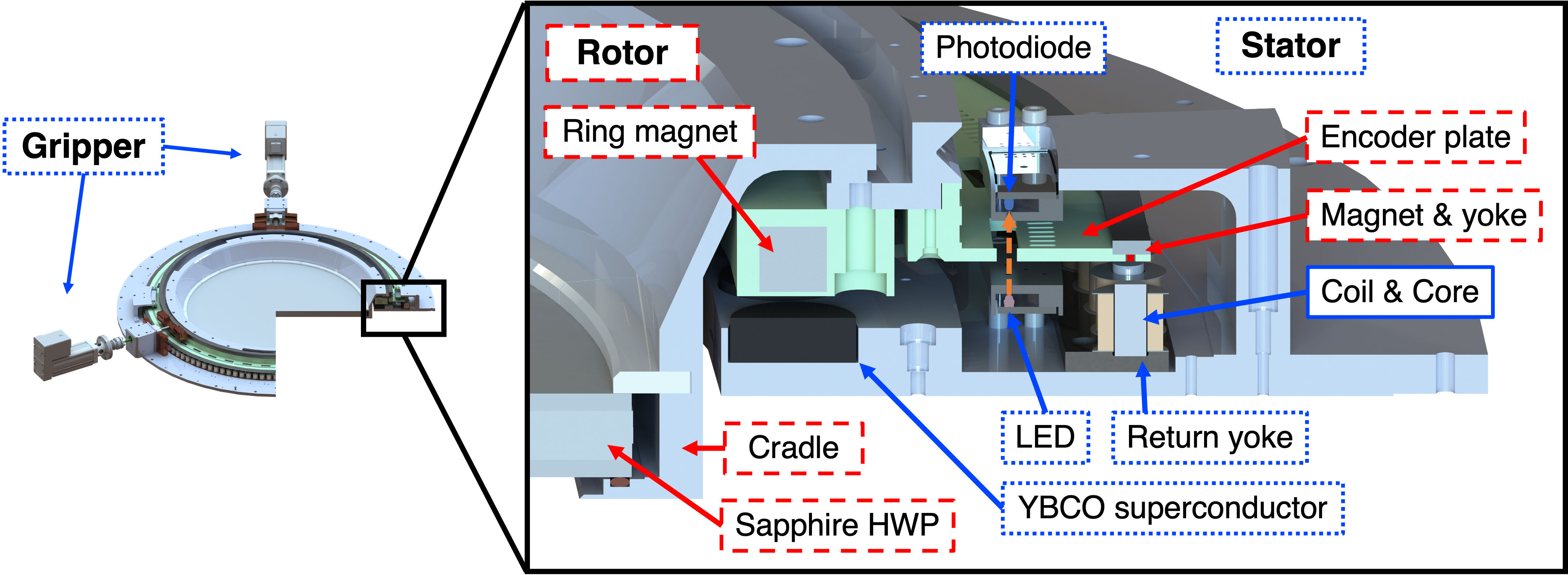}
    \caption{Left panel: CAD views of the CHWP system. Right panel: magnified cross-sectional view of the rotation mechanism. Rotating components are labeled with red dashed boxes, and stationary components are labeled with blue dotted boxes.}
\label{fig:design_cad}
    \end{center}
\end{figure*}

As shown in Figs. \ref{fig:design_cad_sat} and \ref{fig:design_cad}, the non-rotating cryogenic components of the CHWP mechanism are built onto the most skyward flange of the PTC1 stage. 
The main non-rotating components are the superconductor ring, the motor system, and the encoder readheads. Additionally, an aluminum shell on the outer diameter of the CHWP assembly provides a mounting flange for skyward optical components (the PTC1 stage IR blocking filters and related heat strapping) and creates a cold, well-regulated radiative cavity to help control the rotor temperature.

\subsection{Sapphire Stack Mounting}
\label{des-cradle}

\begin{figure}
    \centering
    \includegraphics[width = 0.45\textwidth]{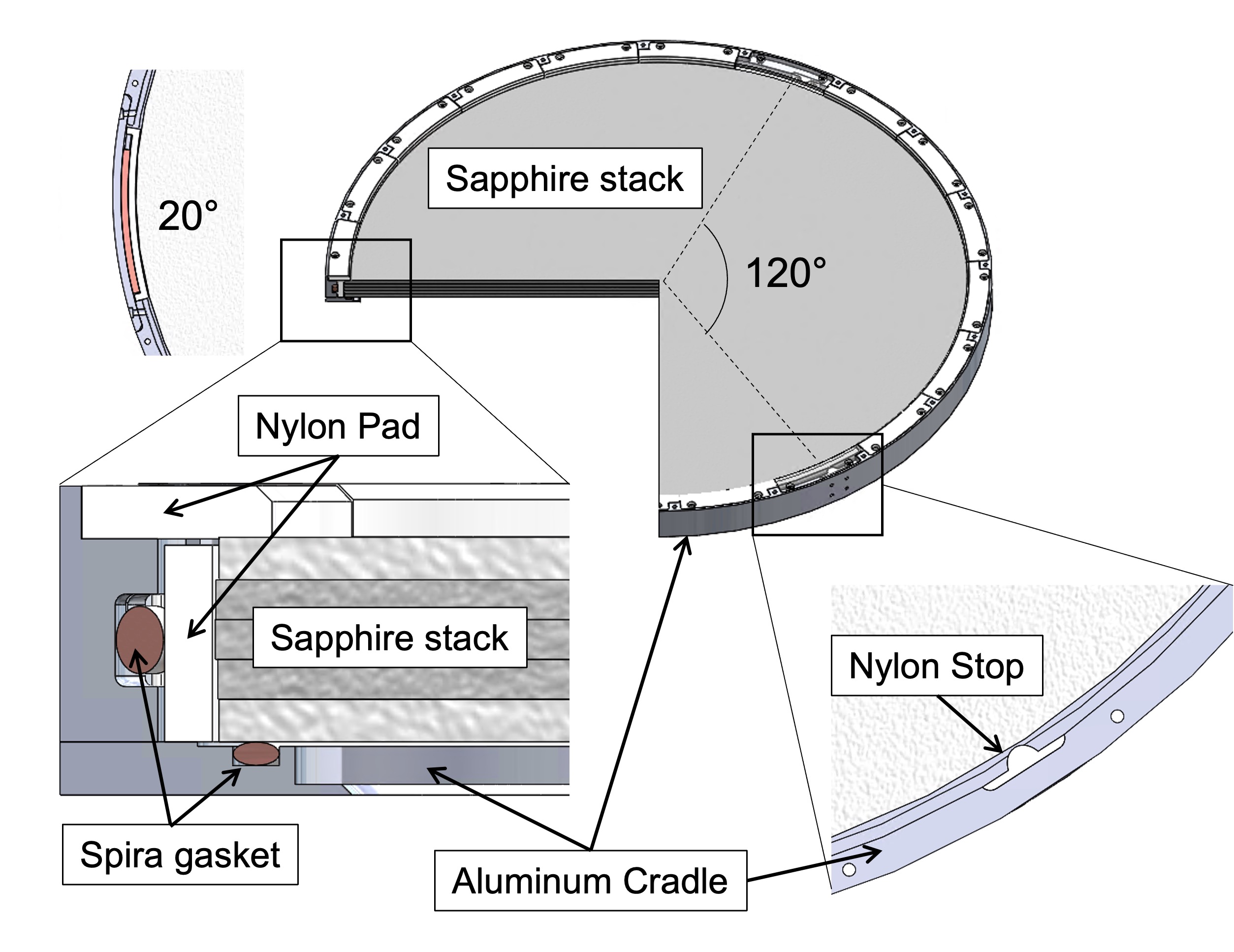}
    \caption{The CAD image of the sapphire stack held in an aluminum cradle. The sapphire stack is held in place vertically by Spira gaskets and nylon pads, and radially by Spira gaskets and nylon stops.}
    \label{fig:cradle_design}
\end{figure}

The CHWP system employs an AHWP with a three-layer-sapphire stack sandwiched by AR layers. Hereafter we refer to this optical element as the sapphire stack. 
It has a $\sim$505~mm diameter, and the total thickness of all five layers is $\sim$20~mm for the MF band. As shown in Fig. \ref{fig:cradle_design}, the sapphire stack is held in an aluminum cradle. 
In the vertical direction, the sapphire stack is held along its circumference between a Spira\footnote{Spira Manufacturing Corp., \protect\url{https://www.spira-emi.com/}} gasket (LS-08) and a series of nylon pads.
Nylon is chosen for its low wear and friction at cryogenic temperatures. \cite{Wisander1966WearAF}
In the radial direction, the sapphire stack is pressed by a 20$^{\circ}$ arc-shaped Spira gasket (SS-16).
On the opposite side, it is held by two nylon stops, each placed at 120$^{\circ}$ from the gasket.
The radial position is designed to be centered after cooldown, taking into account the contractions and the gasket spring constant. 


The cradle is mounted on the rotor baffle, which is shielded from the beam by the stator baffle (Fig. \ref{fig:cradle}).
As discussed in Sec.~\ref{req-opt}, all components except for the sapphire stack, e.g. the cradle aperture and rotor baffle, are designed to intercept the diffracted 90~GHz beam at the $-15$~dB level or below when the rotor is at any position within the mechanical clearance. 
The stator baffle intercepts the diffracted beam at the $-18.5$~dB level at 90~GHz. 

\subsection{Superconducting Magnetic Bearing}
\label{des-smb}
The CHWP employs a 550~mm diameter superconducting magnetic bearing. \cite{Sakurai2020}
While SMBs have been demonstrated in other CHWP systems for CMB observations,\cite{matsumura_phd, klein_cryogenic_2011, johnson_large-diameter_2017, Sakurai2018a, HillPB2bCHWP2020} the SO SMB is the largest of its kind developed to date. 
The SMB is composed of a permanent ring magnet on the rotor and a ring of 61 or 53 disks of yttrium barium copper oxide\footnote{Can Superconductors, CSYL-28 \protect\url{https://www.can-superconductors.com/}} (YBCO) epoxied into an aluminum holder on the stator. 
The number of YBCO disks is different in two versions of SMBs to make the total levitation force similar, because the levitation force per one YBCO disk is different in two versions. 
YBCO is a type II superconductor with transition temperature of $\sim$ 90~K, below which the magnetic flux of the rotor is pinned in flux vortices in the disks, resulting in the loss of all but the rotational degree of freedom.
The ring magnet of 550~mm inner diameter was manufactured by Shin-Etsu Chemical\footnote{Shin-Etsu Chemical Co., Ltd., \protect\url{https://www.shinetsu.co.jp/en/}} and consists of 32 segments of neodymium (NdFeB; N52) fixed in a G10 enclosure. Care was taken in assembly to minimize gaps between the segments in order to maintain an azimuthally symmetric field. 
The number of YBCO disks and the magnet segments that make up the SMB are important design parameters, and making them relatively prime is critical to minimize vibration. 
This optimization of the SMB design and its vibration performance are discussed in Sec~\ref{perf-vib}. 
The friction and stability of this bearing was previously characterized in \citet{Sakurai2020} 

\subsection{Grippers}
\label{des-gripper}
While the YBCO disks are above their transition temperature ($\sim$ 90~K), the rotor is not suspended by the flux-pinning effect and it is necessary to firmly grip it. This is accomplished by a grip-and-release mechanism, referred to hereafter as grippers.
The grippers are a set of three linear actuators \footnote{
SMC Corporation. LEY32C-30B-S11P1} with vacuum adapters manufactured by Huntington. \footnote{Huntington Vacuum Products. \protect\url{https://huntvac.com/}}
The actuator shafts mate to wedged tips mounted on the stator which engage with a matching groove on the rotor when extended. 
This system allows for the precise and repeatable positioning of the rotor with respect to the rigid body of the SAT. The gripper design closely follows the design described in Hill \textit{et~al.},\cite{HillPB2bCHWP2020} with minor modifications. A single gripper can provide a pushing force that is more than twice the weight of the rotor; this is necessary to enable re-gripping and centering at any telescope elevation, as required in Sec. \ref{req-opt}.

The three gripper arms each carry spring-loaded contacts or "touch probes" that align with copper flex-circuit traces on the rotor's outer diameter. These traces connect to a silicon diode thermometer on the rotor.
When the grippers hold the rotor, two spring-loaded contacts touch the flex-circuit traces, measuring the rotor temperature. The measurement is only possible
while the rotor is being held by the grippers.
Thie measurement method can also be used to determine if the rotor is being held firmly. 



\subsection{Motor cryogenic assembly}
\label{des-motor}
\begin{figure}
    \centering
    \includegraphics[width = 0.48\textwidth]{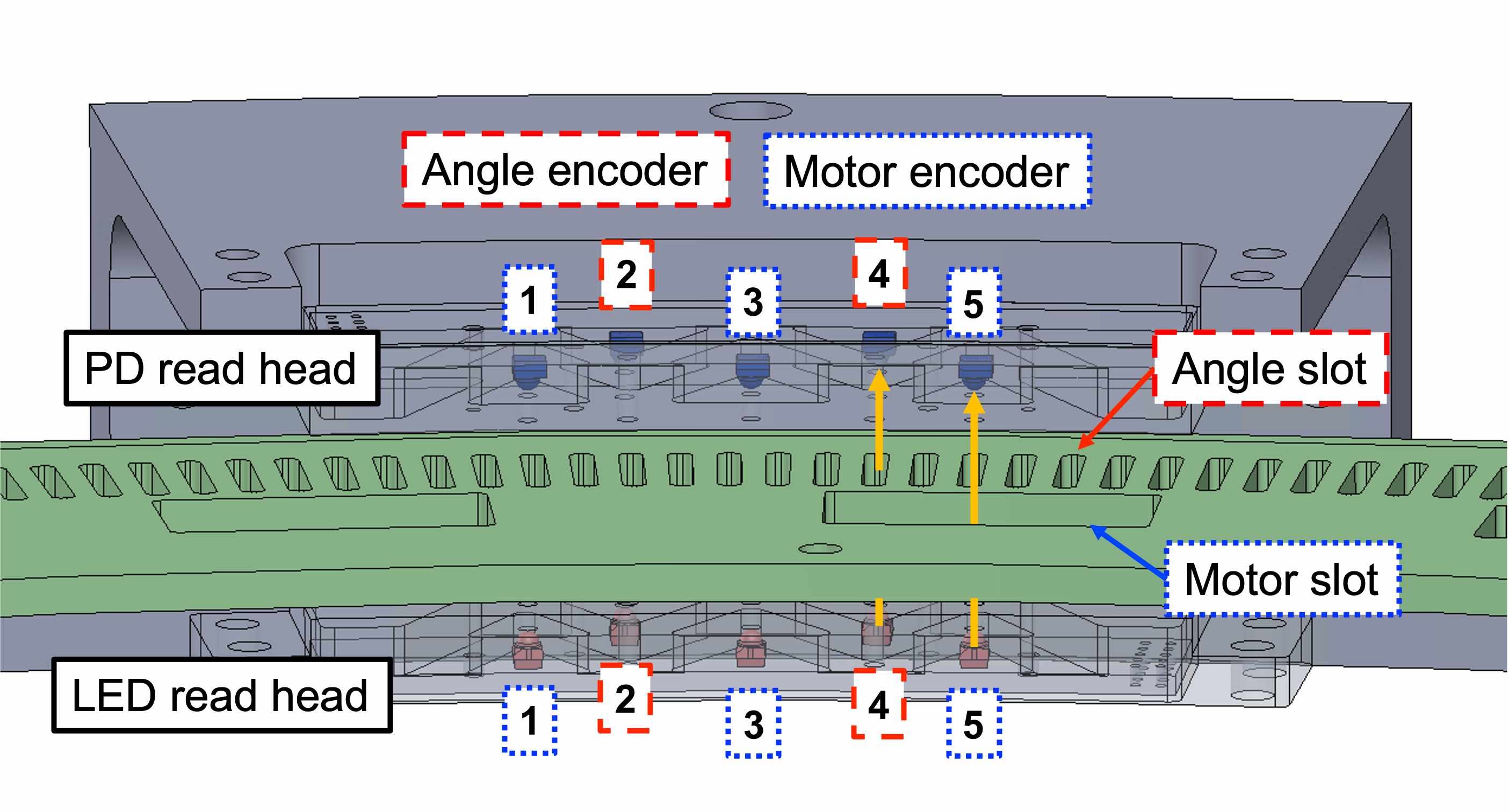}
    \caption{Bird's eye view of the one of the optical encoders and the encoder plate from inside top. The light emitted by the lower LEDs is chopped by the G10 encoder plate, and detected by the PDs hanging over the encoder plate. The two angle encoders chopped by narrower slots are labeled in red, and the three motor encoders chopped by wider slots are labeled in blue.}
    \label{fig:encoder_assembly}
\end{figure}

The motor system that provides torque to the rotor is similar to that described in Hill \textit{et~al.},\cite{HillPB2bCHWP2020} with an increased number of coils due to the larger diameter of the SAT. The magnetic field from the 120 motor coils couples to the 80 magnet sprockets\footnote{K\&J magnetics Inc. D11-N52, \url{https://www.kjmagnetics.com/proddetail.asp?prod=D11-N52}} on the edge of the encoder plate (Fig.~\ref{fig:design_cad}), driving rotation.
We employ two optical encoder assemblies, each with a set of five photo-diodes\footnote{Vishay, TEMD1020, \protect\url{https://www.vishay.com/docs/81564/temd1000.pdf}} mounted on an arm that hangs over a G10 encoder plate on the rotor (Figs. \ref{fig:design_cad} and \ref{fig:encoder_assembly}). 
The LEDs are aligned with a matching set of five IR light-emitting diodes\footnote{Vishay, VSMB294008G, \protect\url{https://www.vishay.com/docs/84228/vsmb294008rg.pdf}} (LEDs) located below the plate. 
The encoder plate is slotted at two different radii in order to chop the light emitted by the LEDs. 
The photo current signal chopped by the 40 wider slots drive the motor coils\footnote{APW electromagnet, FC-6035, \protect\url{https://apwcompany.com/fc-6035/}} by providing feedback through the motor drive electronics (Sec.~\ref{des-driver}), while the signal chopped by the $570 - 1 = 569$ finer slots is fed to the encoder electronics for calculation of the rotation angle of the CHWP (Sec.~\ref{des-daq}). One missing finer slot is a reference for marking the rotor’s absolute rotation angle. 
Two encoder heads are installed at 180$^\circ$ from one another for redundant angle encoding. Data from both encoder heads can be combined to estimate the rotor's off-center displacement, as described in Appendix. \ref{sec:sag-method}.

\subsection{Motor Driver}
\label{des-driver}
The design of the motor drive electronics closely follows that used in \citet{HillPB2bCHWP2020}
Here we highlight the two design improvements.
First, a Proportional-Integral-Differential (PID) feedback system is used further regulate the rotation frequency. For this system, a frequency-to-voltage circuit serves as an input to a PID controller\footnote{Omega, CNI16D54-EIT, \protect\url{https://www.jp.omega.com/pptst/CNI_SERIES.html}} which adjusts the drive voltage and modulates the motor torque. With the PID enabled the frequency control system becomes closed loop and is able to account for unexpected changes to motor efficiency. PID parameters are chosen to minimize long-timescale variations.

Second, a phase compensation circuit is used to further improve rotation stability. 
While the CHWP rotates, the inductance of the motor coils and the counter-electromotive force distort the motor drive current against drive voltage. This distortion acts approximately as an effective phase delay to the motor drive current, and the phase delay increases with the rotation frequency resulting in poor motor efficiency. The phase compensation circuit corrects for this phase delay. Simply compensating for the delayed phase significantly improves the motor efficiency.  
We implement a discrete phase compensation in 60$^\circ$ increments through the sign reversal and the phase swapping of the three-phase motor using relay modules. 
The use of the relay modules eliminates single-point failures and minimizes noise added to the motor coil drive feedback.
Phase compensation with 60$^\circ$ increments is not always optimal, but is easily achieved and provides sufficient control and efficiency. 
We automatically activate the phase compensation circuit when the rotation speed exceeds 1\,Hz. 
The performance of the rotation drive control electronics and the evaluation of rotation efficiency are summarized in Secs.~\ref{perf-mech} and \ref{perf-efficiency}, respectively.


\subsection{Data Acquisition}
\label{des-daq}
Data generated by the CHWP's cryogenic system is routed from the cryostat via four cables to a warm electronics box for processing. 
Acquisition of the rotor angular time stream requires cleaning and digitizing the raw encoder signals sent from the CHWP. This processing follows the method described in \citet{HillPB2bCHWP2020}

The CHWP system also continuously records information from the motor driver, a Hall probe, and four silicon diode cryogenic thermometers.
The diodes are placed on the YBCO superconductor ring, sapphire stack mount on rotor, PTC1 filter plate, PTC2 filter plate, and rotor are continuously monitored by a Lake Shore Model 240. \footnote{Lake Shore, Model 240, \protect\url{https://www.lakeshore.com/products/categories/overview/temperature-products/cryogenic-temperature-modules/240-series-input-modules}} 
A Hall probe\footnote{Lake Shore, HGT-3010 \protect\url{https://shop.lakeshore.com/default/transverse-hall-sensor-3010hgt-3010.html}} 
attached to the YBCO assembly continuously measures the magnetic field intensity around the superconductor (Fig.~\ref{fig:z-sag}). 
The neodymium magnet in the rotor assembly is the primary magnetic source, and its field varies with changes in rotor temperature and displacement,  enabling the Hall probe to monitor both of these properties. \cite{sakurai_estimation_2017}
The monitoring and control of the CHWP system are managed by the Observatory Control System. \cite{OCS2020}

\section{Performance}
\label{sec:performance}
\begin{figure*}
    \centering
    \includegraphics[width = 0.98\textwidth]{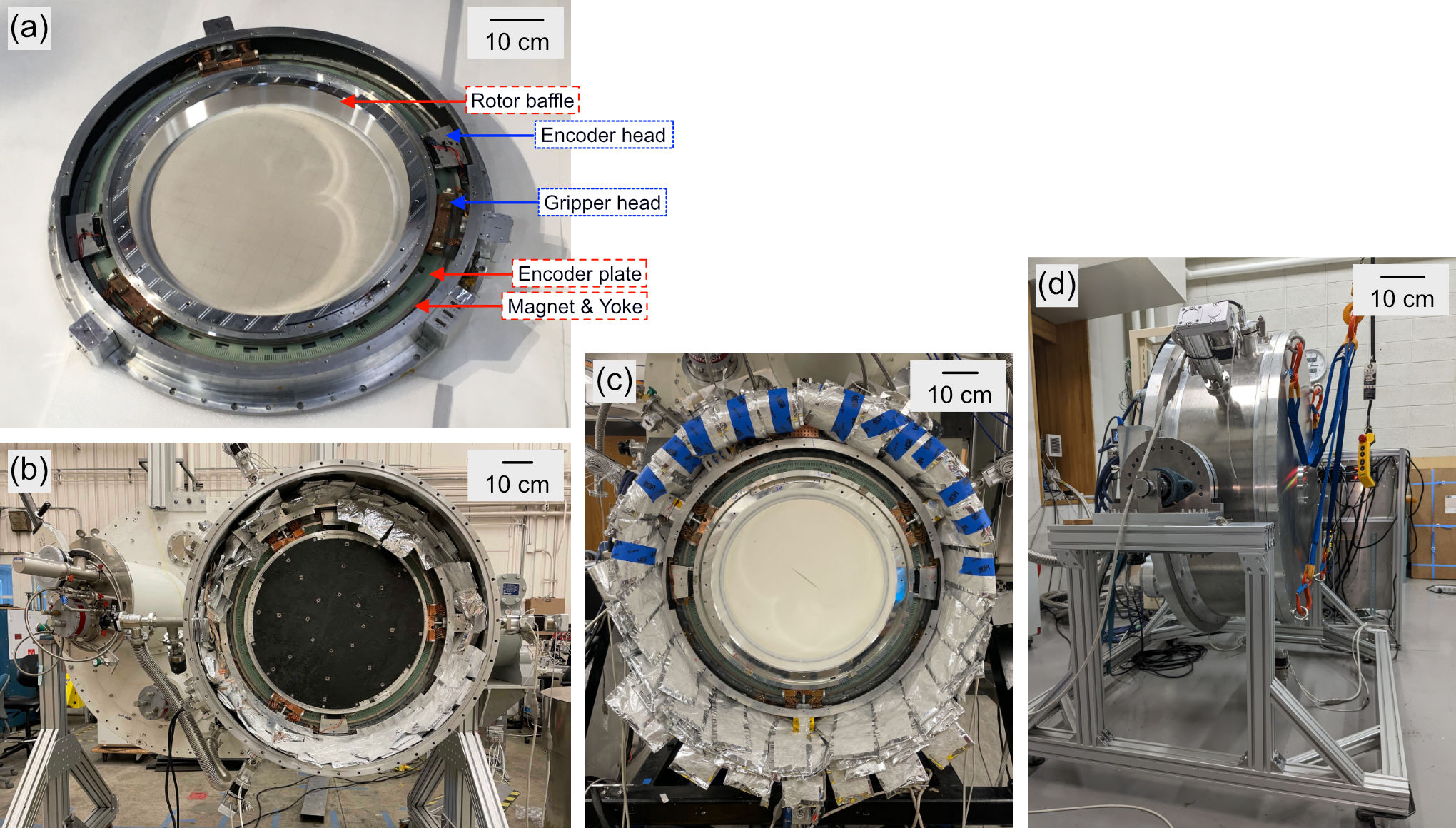}
    \caption{The CHWP rotation mechanisms during testing. Panel (a) shows the rotation mechanism with the sapphire stack for the SAT cryostat at UCSD. Panels (b) and (c) show the rotation mechanisms installed in the SAT cryostat at LBNL and Princeton University respectively. In panel (b), the black object mounted in the rotor is a dummy mass, used to mimic the realistic mechanical and thermal conditions. Panel (d) shows the CWHP test cryostat at the University of Tokyo.}
    \label{fig:lab-testing}
\end{figure*}

We discuss the thermal and mechanical performance of three CHWPs evaluated at the University of Tokyo and in SAT cryostats at the University of California San Diego, Princeton University, and the Lawrence Berkeley National Laboratory (Fig. \ref{fig:lab-testing}). Performance is evaluated using a dummy mass that brings the rotor mass to that expected with the full sapphire stack. The dummy mass is coated with epoxy to mimic the thermal conditions in the relevant IR frequencies. 

\subsection{Thermal}
\label{perf-therm}
\begin{figure}
    \centering
    \includegraphics[width = 0.48\textwidth]{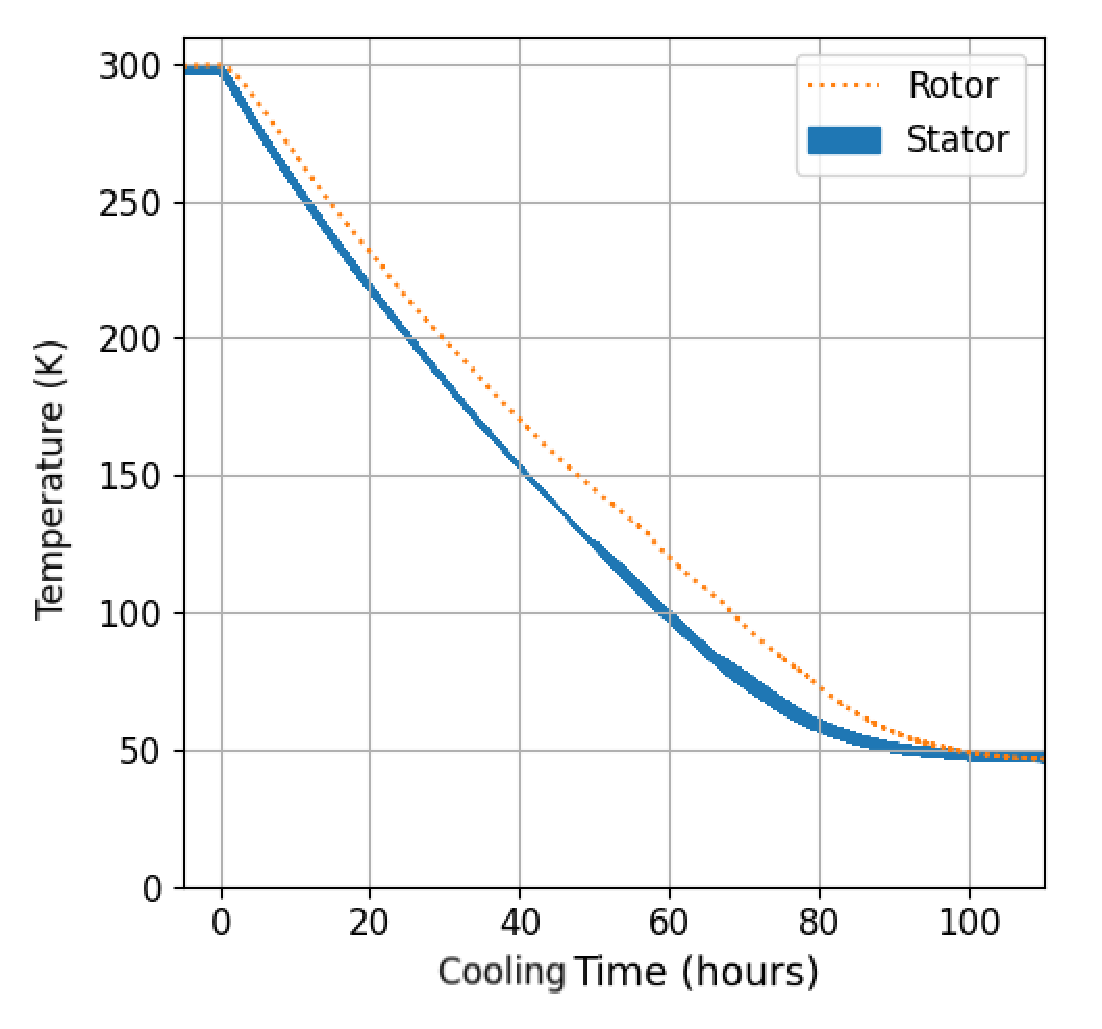}
    \caption{Rotor and stator temperatures during cooldown. The rotor thermalizes within 10 hours of the stator.}
    \label{fig:rotor_thermalization}
\end{figure}

Figure \ref{fig:rotor_thermalization} shows the rotor and stator temperatures during a cooldown. 
Initially, the rotor is held by the grippers and is primarily cooled by conduction through the grippers' copper fingers. The rotor temperature lags behind the stator by 10 hours, well within the 36 hour requirement.

\begin{figure}
    \centering
    \includegraphics[width = 0.48\textwidth]{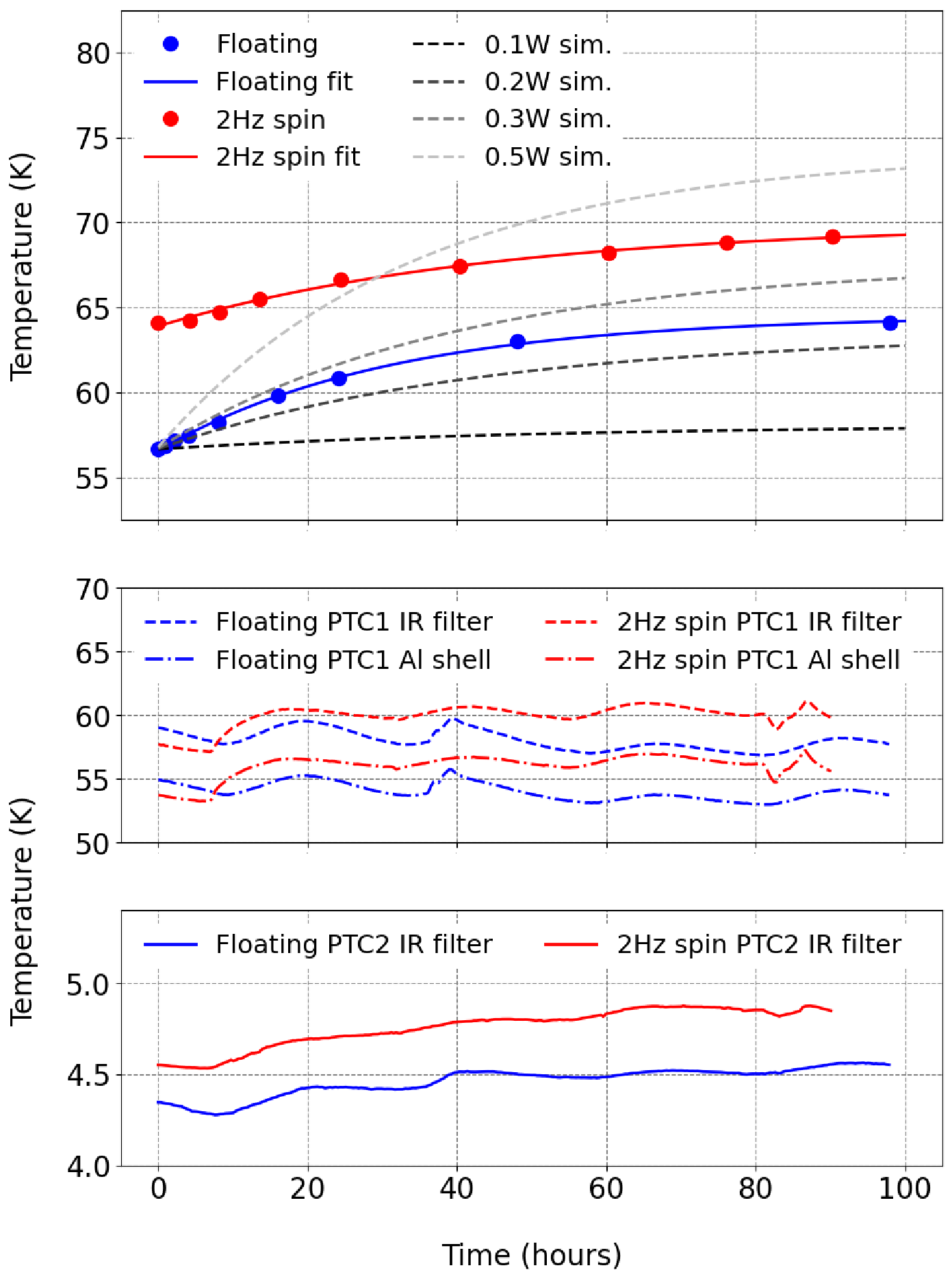}
    \caption{Top panel: The rotor temperature profile while it is floating (blue points) and spinning at 2~Hz (red points), which are fit by Eq.~\ref{eq:rotor_temp_fit} (solid lines). Black dashed lines show the ANSYS thermal simulation results with heat inputs of 0.1, 0.2, 0.3 and 0.5~W, respectively. The initial temperature of the simulation was set to the initial temperature of rotor when released from the grippers. The heat dissipation to the rotor is estimated by comparing the thermal equilibrium temperature between the fit result and the simulation. Middle panel: Colored dashed lines represent temperatures of the alumina IR filter and aluminum shell.
    Bottom panel: PTC2 IR filter temperature profile while the rotor is floating (blue) and spinning at 2~Hz (red).}
    \label{fig:thermal_profile}
\end{figure}

Once the YBCO disks become superconducting and the rotor temperature stabilizes, the rotor is ungripped and begins to levitate. 
Figure \ref{fig:thermal_profile} shows the rotor temperature as a function of time, while it is floating (blue points) and when it spins at 2~Hz (red points) after temperature stabilization.
In order to collect temperature data with the rotor spinning, its rotation is intermittently stopped and its temperature is measured using the gripper touch probes (Sec.~\ref{des-gripper}). 
This process is performed once every few hours with the duration of each touch not exceeding 60 seconds, leading to little disturbance in the rotor’s temperature trend.
The blue points inform us of optical loading where the IR loading is dominant, 
while the red points provide estimates of the excess loading generated by rotation.
Each set of points are fit by 
\begin{equation}
    T(t) = T_{e} + \Delta T \exp({-t}/{\tau}),
    \label{eq:rotor_temp_fit}
\end{equation}
where $T_{e}$ represents the estimated rotor temperature at steady state, 
$\tau$ is the time constant, and $\Delta T$ is the difference between the initial and steady state temperature.
The steady state temperature of the CHWP rotating at 2 Hz is 70 K, which satisfies the requirement of $\leq$ 85~K. 
The time constants of the blue and red curves are 31~hours and 46~hours, respectively.

\begin{figure}
    \centering
    \includegraphics[width = 0.48\textwidth]{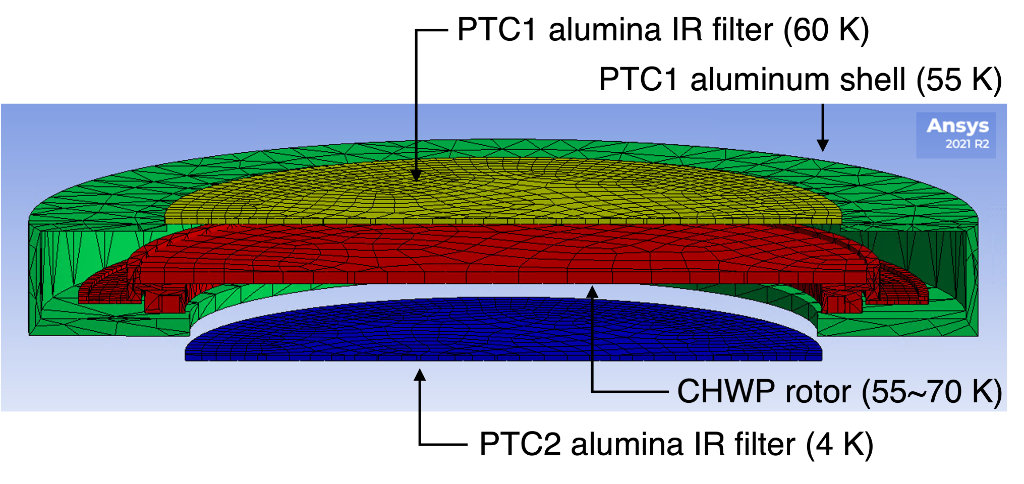}
    \caption{The ANSYS thermal model used to simulate the amount of thermal loading from the CHWP rotor to the PTC2 stage. Temperatures of the PTC1 aluminum shell, the PTC1 and PTC2 alumina IR filters are set at 55~K, 60~K and 4~K respectively. The rotor temperature is varied from 55~K to 70~K to estimate the thermal loading on the PTC2 alumina IR filter. Image used courtesy of ANSYS, Inc.}
    \label{fig:ansys}
\end{figure}

Figure \ref{fig:ansys} shows the model for an ANSYS thermal simulation, which is used to further characterize the CHWP thermal cavity as well as the heat input to the rotor. 
For the simulation we simply model the floating rotor surrounded by the PTC1 aluminum shell and the PTC1 and PTC2 alumina IR filters.
Based on our measurements, we set the temperatures of aluminum shell and alumina filters in the simulation to 55~K, 60~K and 4~K, respectively. 
The infrared emissivity of the alumina IR filter is set to 0.8,\cite{alumina_filter} while that of the aluminum shell is set to 0.96. Here the inside of the aluminum shell is covered with an infrared black body. \cite{berkeley_black}
Since there is no physical contact, the rotor exchanges heat with other components only by thermal radiation.
Figure~\ref{fig:thermal_profile} shows the simulated time series of the rotor temperature with a constant heat input to the floating rotor. 
By comparing the data with the simulation result, the heat input to the rotor is estimated to be 220~mW when the rotor is floating, and 390~mW when the rotor is rotating at 2~Hz. 
The additional thermal loading on the PTC2 IR filter due to the presence of the rotor is estimated to be 120~mW, which is sufficiently small compared to the cooling capacity of the PTC2 stage, 1.8~W. 

\subsection{Rotation control}
\label{perf-mech}
\begin{figure}
    \centering
    \includegraphics[width=0.48\textwidth]{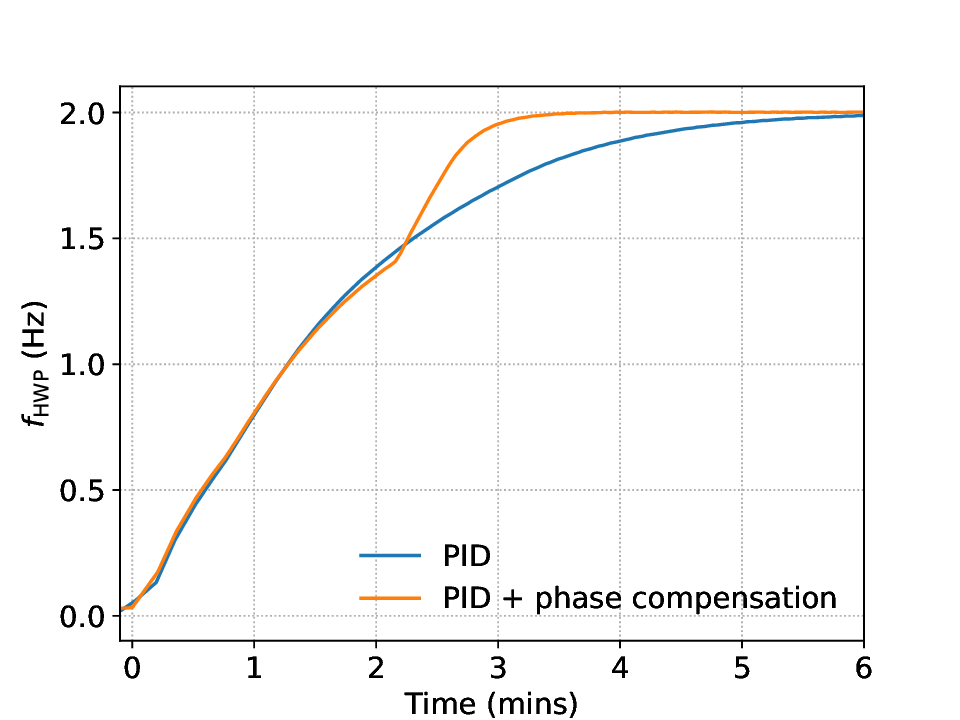}
    \caption{The spin up curves. The blue curve shows the spin up curve with PID control. The rotation frequency stabilizes in $\sim$ 7 minutes. The orange curve shows the spin up curve with phase compensation activated at 2.2 minutes. With phase compensation, the rotation frequency stabilizes in $\sim$ 4 minutes.}
    \label{fig:spinup}
\end{figure}

\begin{figure}
    \centering
    \includegraphics[width=0.48\textwidth]{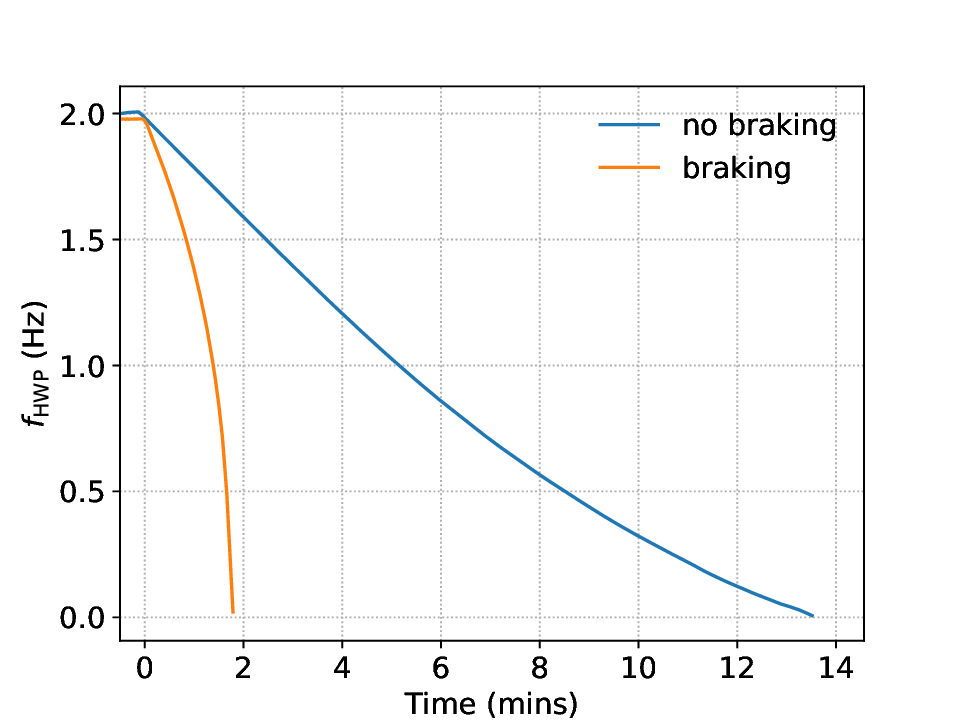}
    \caption{The spin down curves. The blue curve shows the spin-down curve of HWP without braking. Braking can bring the HWP to a stop in less than 2 minutes (orange curve).}
    \label{fig:spindown}
\end{figure}

Figures \ref{fig:spinup} and \ref{fig:spindown} show the performance of the PID control and the phase compensation described in Sec. \ref{des-driver}.
When the rotation PID feedback is on, the rotor accelerates to a steady rotation at 2~Hz within 7 minutes, and decelerates down to 0~Hz within 3 minutes. When phase compensation is used in conjunction with the PID, the spin-up time to steady rotation at 2~Hz is reduced to less than 4 minutes. 

\begin{figure}
    \centering
    \includegraphics[width=0.48\textwidth]{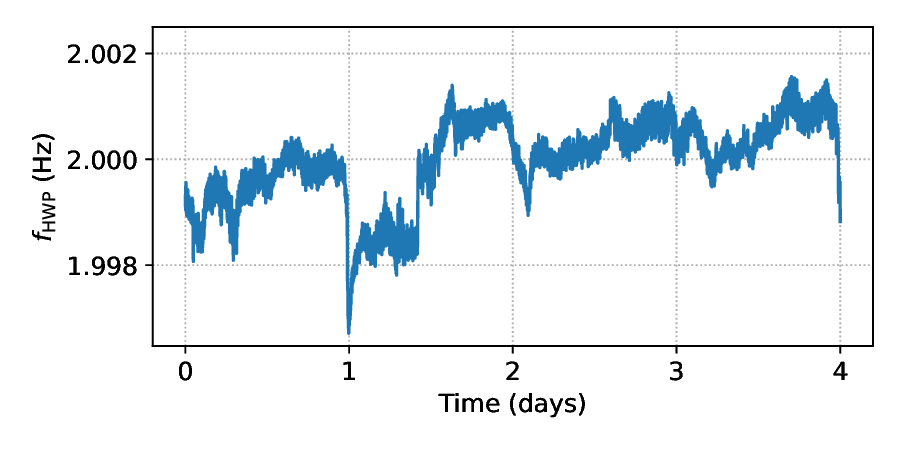}
    \caption{Demonstration of the rotational stability of the CHWP at 2~Hz over 4~days using PID control.}
    \label{fig:stability}
\end{figure}

Figure \ref{fig:stability} demonstrates the stability of the CHWP's rotation at 2~Hz over 4~days using PID control. 
The achieved stability of $\pm$ 5~mHz is well within the required $\pm$ 10~mHz. 
Although not tested in-lab, the stability of the PID control loop over even longer timescales is expected to remain below the requirement.
We additionally performed an in-lab azimuth scan to demonstrate stable operation of the CHWP in observation-like conditions. The receiver was held at a constant elevation angle of 50$^\circ$ and subject to an angular throw of 12$^\circ$, constant scan velocity of 1$^\circ$/sec, and turnaround acceleration of 1$^\circ$/sec$^2$. While the nominal SAT angular throw will be wider, more frequent turnarounds during this test enabled an evaluation of scanning effects on PID performance under more strenuous operating conditions.

\begin{figure}
    \centering
    \includegraphics[width = 0.48\textwidth]{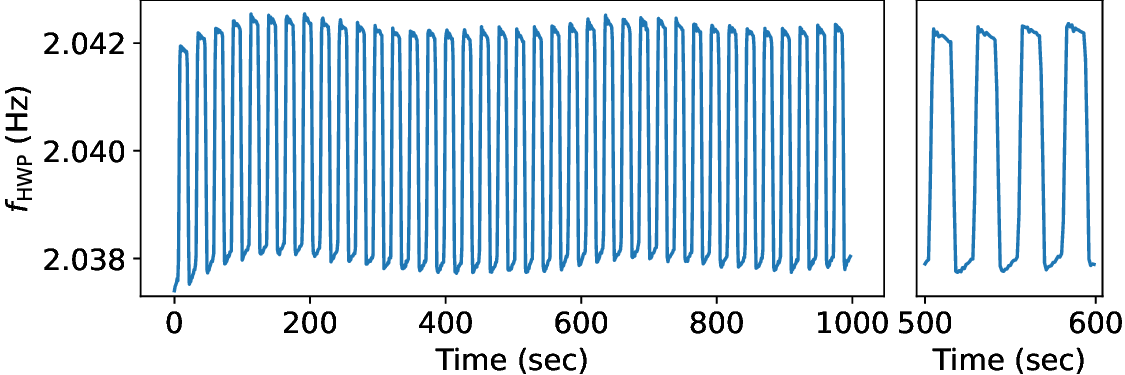}
    \caption{Time stream of the measured HWP rotation frequency during the constant elevation scan of the SAT receiver at an elevation of 50$^\circ$. The long timescale PID rotation control is enabled.}
    \label{fig:scan_hwp}
\end{figure}

Figure \ref{fig:scan_hwp} shows a time stream of the measured rotation frequency during scanning where we observe a scan-synchronous modulation of the measured frequency ($\pm 2$~mHz). 
This is due to the fact that the encoders and SAT detectors are fixed in the telescope reference frame, the magnetically levitated CHWP is not. 
Thus the measured modulation arises from conservation of angular momentum of the CHWP about its rotation axis. 
The angular velocity of the CHWP during scanning is given by the following, the detailed derivation of which is described in Appendix~\ref{sec:scan-mod}:
\begin{align}
    \label{eq:scan_modulation}
    \dot{\chi}(t) = \bar{\dot{\chi}} - \dot{\phi}(t)\sin(\theta_\mathrm{el}),
\end{align}
where $\phi$ is the azimuth angle of the telescope, $\theta_\mathrm{el}$ is the elevation angle of the telescope and $\bar{\dot{\chi}}$ is the average angular velocity of the CHWP.
The PID is tuned for longer-timescale stability and reacts little to the azimuthal scan modulation. This demonstrates that the accuracy of the encoding system and the stability of rotation frequency are well within the requirement including the effect of scan modulation.

\subsection{Rotation efficiency}
\label{perf-efficiency}
\begin{figure}
    \centering
    \includegraphics[width = 0.48\textwidth]{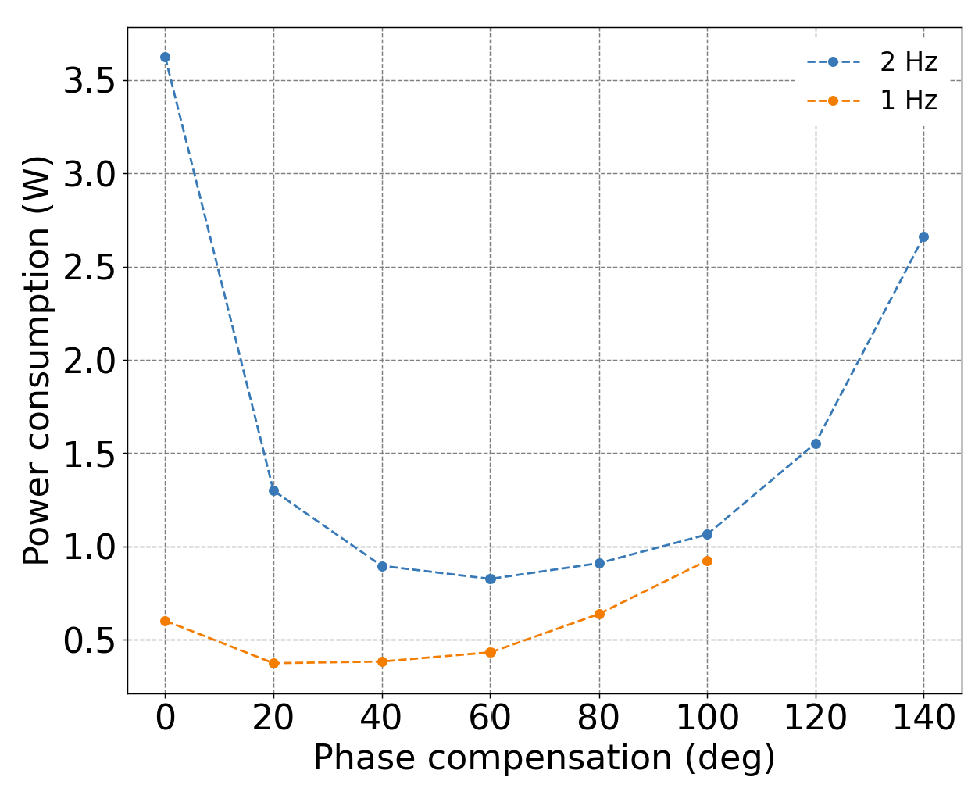}
    \caption{The power consumption of the voltage source of the three-phase motor as a function of phase compensation angle. The blue (orange) line is for the rotor rotating at 2 (1)\,Hz. A digital phase compensation circuit implemented by a microcontroller is used to apply an arbitrary amount of phase compensation and to explore the motor efficiency as a function of the compensated phase. We find that the near-optimal compensation angle is 60$^\circ$, which can be achieved by discrete phase compensation.
    As such, we adopt the simpler and more robust discrete phase compensation in the actual implementation. The digital phase compensation circuit is used only for this investigation. 
    }
    \label{fig:dissipation}
\end{figure}

Table \ref{tab:dissipation} summarizes the thermal dissipation of the rotation mechanism to the stator and PTC1 stage.
There are four primary sources of thermal dissipation from the CHWP:
a) Joule heating of the driving coils, where the phase compensation of the motor driver (Sec.~\ref{des-driver}) plays an important role, b) hysteresis loss in the SMB, c) eddy current loss in the SMB, and d) dissipation from the optical encoders.
a) is characterized by the impedance and the bias voltage of the motor coils, while b) and c) are characterized by fitting the spin-down curve, using the method described in \citet{Sakurai2020}
As noted in Table~\ref{tab:dissipation}, b) and c) are dependent on the elevation angle, or the gravity vector direction. Lower elevation angles lead to a larger distance between the YBCO and the magnet ring of the rotor along the optical axis, resulting in a weaker and smoother magnetic field coupled to the YBCO, thus diminishing the loss from hysteresis and eddy currents. 
The total dissipation from the motor and SMB (a+b+c) is measured by comparing the loading on the PTC1 stage when the CHWP is operating and when it is not.
The power dissipation of the optical encoders is mainly from the LEDs, which is estimated to be smaller than 1~W based on the bias current and voltage.

a) is reduced by optimizing the phase compensation angle of the drive motor (Sec.~\ref{des-driver}). 
The phase compensation angle was optimized using the power consumption of the motor driving voltage source,\footnote{Kikusui electronics corp. PMX35-3A} which includes dissipation both inside and outside of the receiver.
Figure~\ref{fig:dissipation} shows the power consumed by the voltage source as a function of the phase compensation angle of the drive motor. 
The power consumption with a 2-Hz rotation achieves minimum at 60$^\circ$. 
With a motor drive phase compensation of 60$^\circ$, the power consumption is reduced from 3.6$\pm$1.0~W to 0.8$\pm$0.1~W, and the dissipation of the motor on the stator is reduced from 3.1$\pm$1.0~W to 0.5$\pm$0.1~W. 
This results in a total power dissipation of $\leq 1.6$~W summing motor dissipation with that of the LEDs, which is below the requirement of 3~W. 
\begin{table}
\caption{Dissipation of rotation mechanism on stator, PTC1~stage.}
\label{tab:dissipation}
\begin{ruledtabular}
\begin{tabular}{llc}
Motor & Joule heat of driving coils & 0.4$\pm$0.1\,W (2.6$\pm$1.0\,W)\footnotemark[1] \\
SMB & Hysteresis loss of SMB & 0.083 W/ 0.084 W\footnotemark[2]\\
& Eddy current loss of SMB & 0.058 W/ 0.099 W\footnotemark[2]\\
& Total & 0.5$\pm$0.1~W (3.1$\pm$1.0~W)\footnotemark[1] \\ \hline
Encoder & Power dissipated by LEDs & $\leq$ 1~W\footnotemark[3] \\
\end{tabular}
\footnotetext[1]{Dissipation when phase compensation is activated (not activated). The error bar is the systematic variation due to due to its dependence on the elevation angle and the rotation direction.}
\footnotetext[2]{Dissipation at an elevation of 50$^\circ$~/~90$^\circ$.}
\footnotetext[3]{10 LEDs are biased with 50~mA and 2~V.}
\end{ruledtabular}

\end{table}

As we discuss in Sec.~\ref{perf-sag}, the elevation-dependent off-center displacement of the rotor induces a phase shift between the measurements of the two encoders. Because the motor drive voltage source uses feedback from the encoder to regulate its output (Sec.~\ref{des-driver}), the rotation efficiency is dependent on this elevation-dependent phase shift and also on the rotation direction, resulting in the systematic variation of the Joule heat of the driving coils and the power consumption by the voltage source. 
As we see from Fig.~\ref{fig:dissipation}, the rotation efficiency is sensitive to the phase shift of the motor when phase compensation is not activated, but is insensitive when 60$^\circ$ of phase compensation is activated. 
Therefore, the phase compensation enables to achieve robust rotational efficiency regardless of changes to the off-center displacement and rotation direction.
The systematic variation in the thermal dissipation of the rotation mechanism is negligible compared to the elevation-dependence of the PTC's cooling capacity.\cite{TSAN2021103323}

\subsection{Vibration} \label{perf-vib}
The characteristic vibration frequencies of the SMB are key parameters for the vibrational performance of the CHWP. From the mass of the rotor and the displacement we evaluate in Sec.~\ref{perf-sag}, the spring constants of the SMB parallel and perpendicular to the optical axis are estimated to be 9.1(17)$\times10^4$~N/m and 5.2(8.4)$\times10^4$~N/m respectively for 48(61) YBCO tiles. 
From these values, we determine the corresponding characteristic vibrational frequencies to be 9(13)~Hz and 7(9)~Hz respectively, which are in agreement with the results of \textcite{Sakurai2020}

The primary instrumental parameter that impacts the vibrational performance is the number of YBCO disks of the SMB. 
Throughout our evaluation of the SAT’s susceptibility to CHWP induced vibrations,
our design of the SMB evolved from 48 disks to 53 or 61. This section presents their comparison.
We employ two methods to characterize vibration, one by using a three-axis accelerometer \footnote{Analog Devices, adxl345 \protect\url{https://www.analog.com/en/products/adxl345.html}} mounted on the cryostat near the rotation mechanism, and the other by measuring the temperature increase of the 100-mK focal plane stage while sweeping through CHWP rotation frequencies.

\begin{figure*}
    \centering
    \includegraphics[width = 0.95\textwidth]{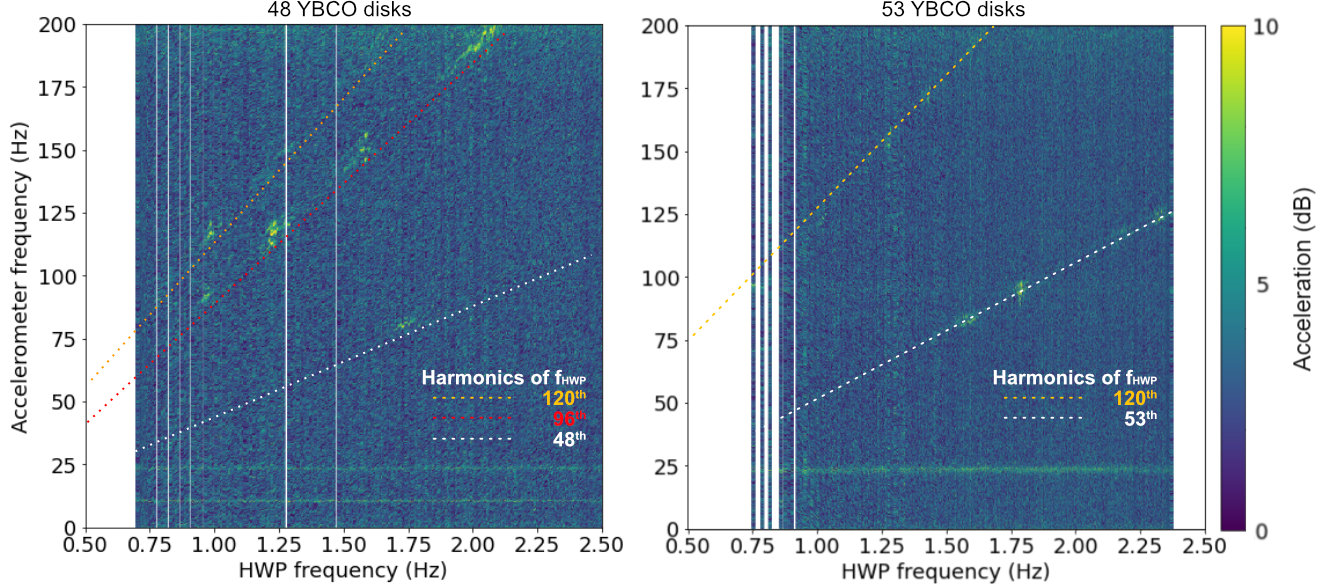}
    \caption{Comparison of the vibration of the SMB with different numbers of YBCO disks. The color scale is normalized by the noise level of the accelerometer. 
    Vibration is observed at $f_\mathrm{HWP}$ multiplied by an integer times a common multiple of the characteristic numbers of the SMB: the number of ring magnet segments (=32), the number of YBCO disks (=48, 53), or the number of coils (=120). The vibration observed on the 48-disk SMB at $f_\mathrm{HWP}$ multiplied by the least common multiple of the number of YBCO disks and ring magnets segments (=96) is not observed on the 53-disk SMB.}
    \label{fig:utokyo_vib}
\end{figure*}

Figure \ref{fig:utokyo_vib} shows the spectrogram of the vibration parallel to the optical axis. 
We do not observe significant vibration perpendicular to the optical axis.
Vibrations were measured at frequencies equal to $f_\mathrm{HWP}$ multiplied by the number of coils (= 120), or the number of YBCO disks (= 48, 53) multiplied by an integer.
In the 48-disk setup, the 96th harmonic of $f_\mathrm{HWP}$ was the largest vibrational mode.
96 is the least common multiple of the number of YBCO disks (= 48) and the number of NdFeB magnet segments (= 32), and this relationship yielded vibrational coupling that produced higher amplitude vibration.
In the 53-disk setup, the number of YBCO disks and NdFeB magnets are relatively prime (do not have common divisors larger than 1) and we did not observe vibrational mode associated with their coupling. 

\begin{figure}
    \centering
    \includegraphics[width = 0.48\textwidth]{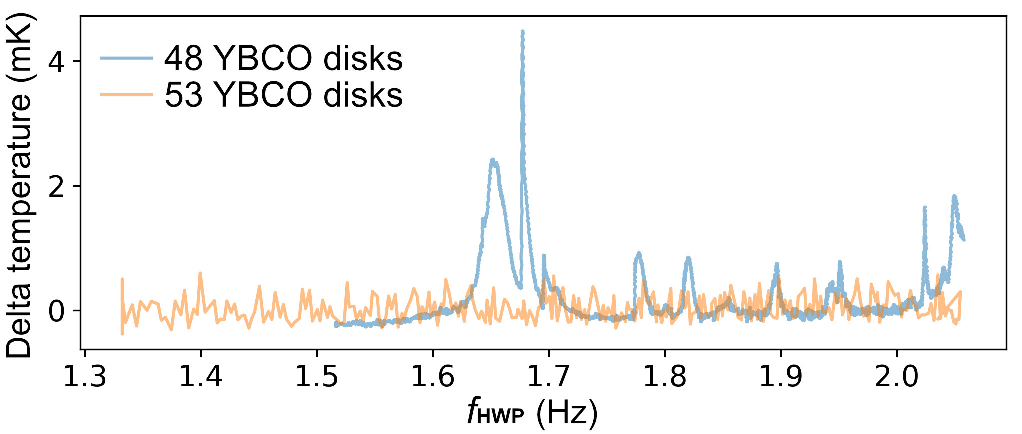}
    \caption{Comparison of the effect of vibration on the focal plane temperature in two SMBs with different number of YBCO disks. The nominal focal plane temperature is 100\,mK.
    }
    \label{fig:fp_warmup}
\end{figure}

The influence of vibration on the focal plane temperature in two representative SMBs with different numbers of YBCO disks is shown in Fig.~\ref{fig:fp_warmup}.
The rotation of CHWP was monotonically swept in frequency from 0.5 to 2.3~Hz over a 12 hour duration while measuring the temperature of the cryogenic stages.
Since these two SMBs are tested in SATs with different instrumentation, a direct comparison of the thermal sensor noise is not possible, however the resonant heating response observed in a 48-disk setup is not seen in a 53-disk setup.

\subsection{Rotor alignment and displacement} \label{perf-sag}
There are two degrees of freedom that we consider important for the alignment of the CHWP rotation mechanism: alignment along the optical axis, involving displacement perpendicular to the plane of the SMB, and center alignment, involving displacement along the SMB plane. In this section we discuss alignment performance along these degrees of freedom, in addition to the effect of temperature on alignment. 
The quality of the rotor alignment is determined both by the initial centering established by the grippers and by the stiffness of the SMB. 

Because the magnetic flux-pinning acts as a restoring force with a finite spring constant, 
when the rotor is released from the grippers it displaces due to gravity, finding a new equilibrium position that  depends on the elevation angle. 
Due to its azimuthal symmetry, the SMB is most stiff along the optical axis. \cite{klein_cryogenic_2011}
The off-center displacement produces a phase shift between the two angle encoders, resulting in a difference between the calculated and actual rotational angle and the encoded angle. 
Therefore, the evaluation of off-center displacement is particularly important. 

The stiffness of the SMB is the product of the critical density current of the YBCO, the magnetization of the NdFeB permanent magnet ring, and a constant determined by the geometry of the SMB. 
A temperature increase from 50~K to 85~K results in $\sim$90\% reduction in the critical density current\cite{Tiwari1996} and $\sim$10\% reduction in the magnetization.\cite{mag_thermometer}
Thus, by measuring the temperature dependence of the rotor displacement we can determine the temperature dependence of the stiffness and obtain the maximum operating temperature of the SMB.

Each of the three primary CHWP alignment considerations can be evaluated independently. 
Alignment along the optical axis is characterized at 50~K with the methods described in Appendix~\ref{sec:sag-method}. The largest displacement occurs at an elevation angle of 90$^\circ$, resulting in a displacement of $2.0\pm0.3$~mm, which is well within the designed clearance of 5 mm.

Center alignment is evaluated using the method described in Appendix~\ref{sec:sag-method}. Figure \ref{fig:sag_vs_elevation} shows the measured off-center displacement at different elevation angles and for SMBs with different numbers of YBCO disks. The off-center displacement was measured seven times at an elevation of 90$^\circ$, and the initial alignment accuracy was found to be $\leq$ 1~mm. 
Inaccurate initial centering by the grippers while cooling through the YBCO transition produces a non-zero displacement at an elevation of 90$^\circ$.
As Fig.~\ref{fig:sag_vs_elevation} shows, the off-center displacement decreases as the number of YBCO disks is increased, and is smaller than 3.5 mm with 61 YBCO disks at elevation angles larger than 20$^\circ$.
\footnote{The minimum elevation angle of the SAT platform is 20$^\circ$.}
This satisfies the requirement for optical and mechanical clearance of $\leq$ 5~mm even when combined with the initial alignment accuracy $\leq$ 1 mm.
\begin{figure}
    \centering
    \includegraphics[width = 0.48\textwidth]{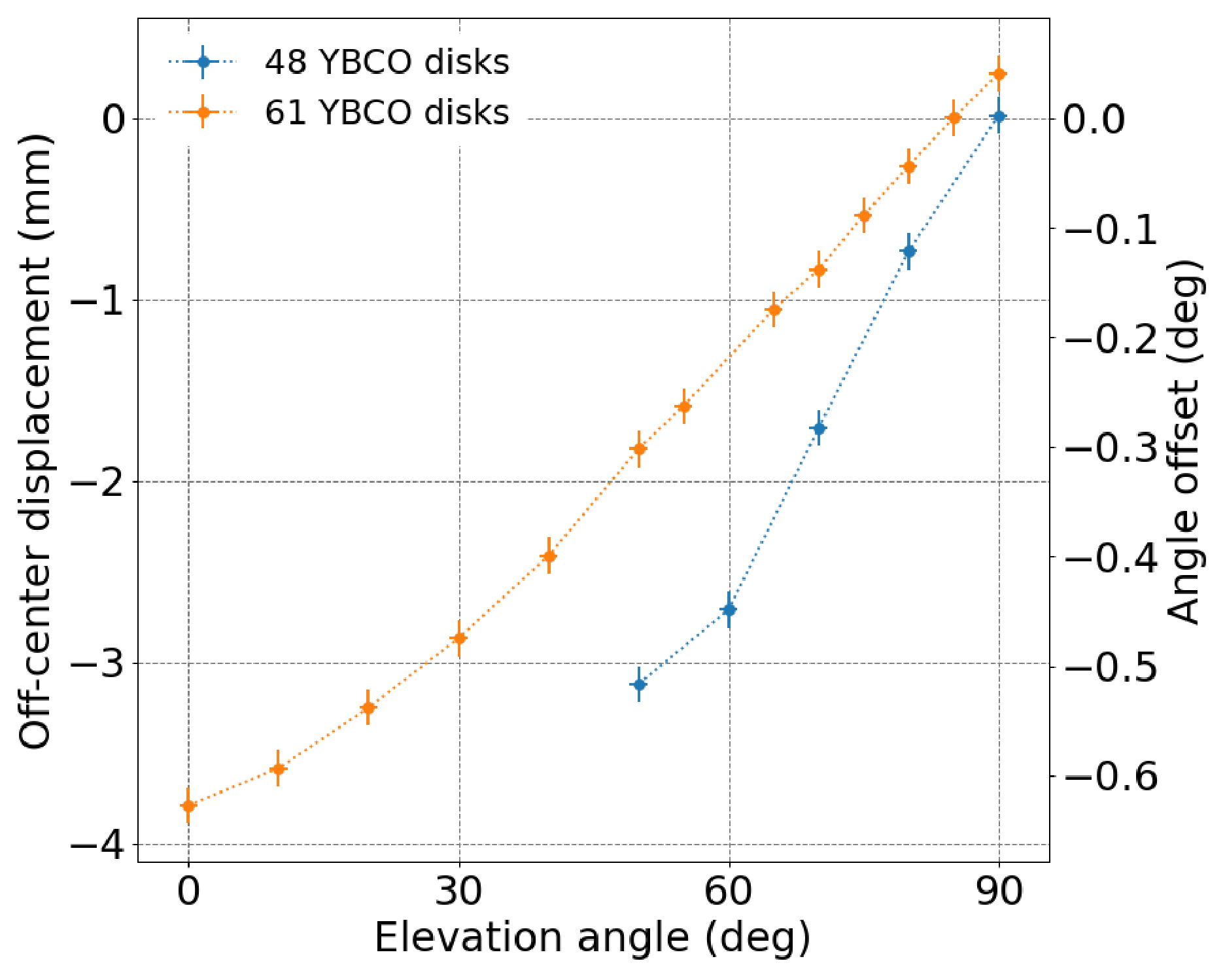}
    \caption{The off-center displacement of the rotor at different elevation angles and different number of YBCO disks. The elevation angle of 90$^\circ$ refers to pointing at zenith (Fig.~\ref{fig:scan_mod_sketch}). The angle offset is an additional measured angle difference between the two encoders due to the off center displacement of the rotor (Eq.~\ref{eq:phasediff}) }
    \label{fig:sag_vs_elevation}
\end{figure}

\begin{figure}
    \centering
    \includegraphics[width = 0.48\textwidth]{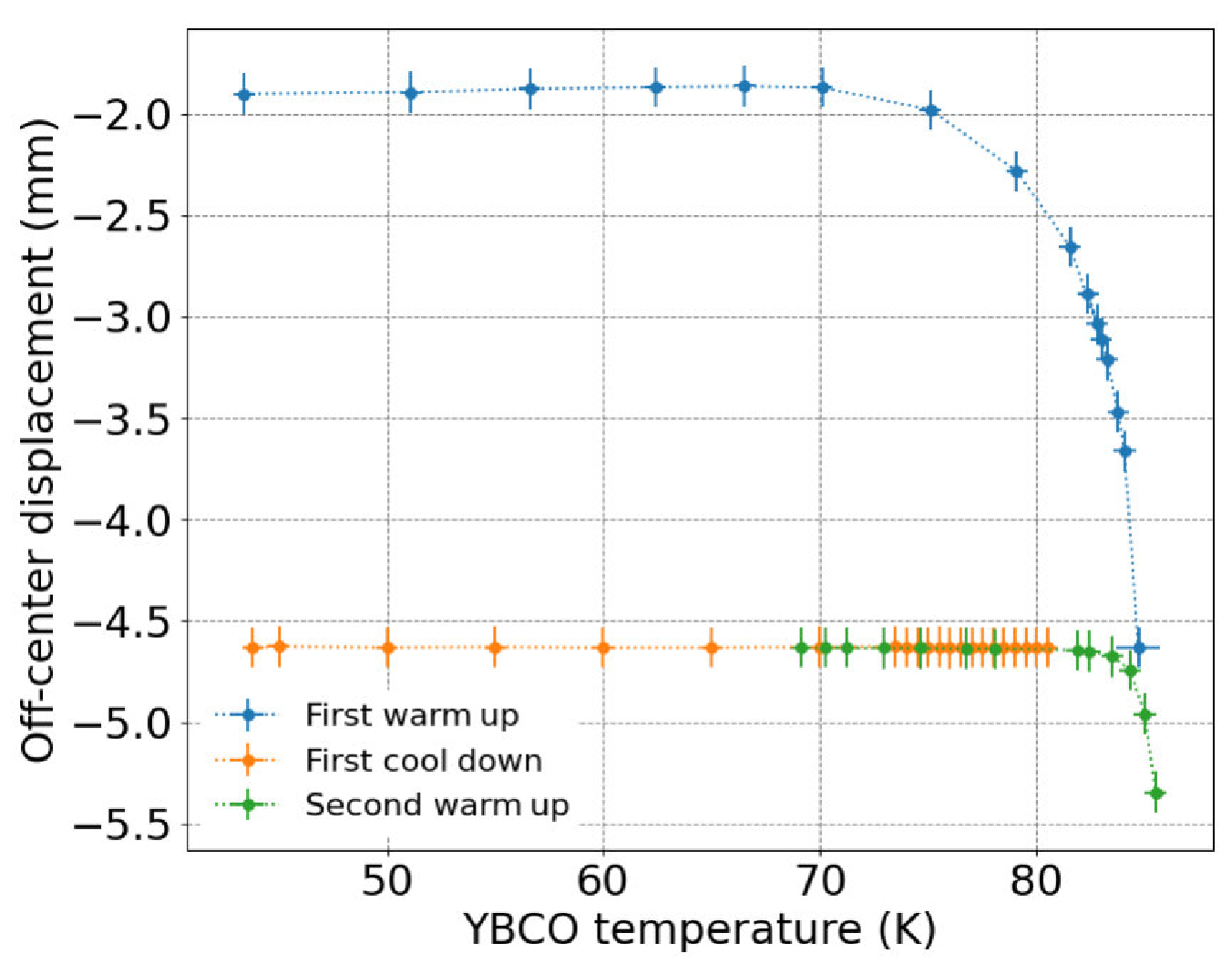}
    \caption{The off-center displacement when the YBCO temperature is increased and decreased to evaluate the maximum operating temperature of the SMB. During the measurements, the rotor is continuously rotating at 2~Hz and the elevation angle is 50$^\circ$.}
    \label{fig:sag_vs_temperature}
\end{figure}

Finally, the temperature dependence of the displacement (Fig.~\ref{fig:sag_vs_temperature}) is evaluated to determine the maximum operating temperature through the following three steps:
\begin{enumerate}
    \setlength{\itemsep}{0pt}
    \setlength{\parskip}{0pt}
    \setlength{\parsep}{0pt}
    \item The temperature of the YBCO is gradually increased to 85~K. The off-center displacement is constant up to 70~K, but rapidly increases after reaching 70~K. 
    \item The YBCO temperature is gradually lowered from 85~K. The off-center displacement does not change throughout this process. 
    \item The YBCO temperature is gradually increased to observe where the off-center displacement starts to increase again. The off-center displacement remains the same up to $\sim$ 85~K, but begins to rapidly increase when the temperature rises above 85~K.
\end{enumerate}
With this test, we find that the maximum operating temperature of the SMB is not the transition temperature of the YBCO ($\sim$95~K) but 70 K. This temperature requirement is satisfied with sufficient margin under the normal operating conditions of our system.
We additionally find that, if the off-center displacement occurs due to the temperature increase, it cannot be restored by simply lowering the temperature of the YBCO. 
Once the displacement has occurred, it is retained unless the SMB is brought to a higher temperature.
In order to re-center the rotor, it is necessary to grip it at the center and warm up the YBCO disks above their transition temperature, followed by a re-cooling to flux-pin the rotor in the correct position.

\subsection{Angle encoding accuracy}
\label{perf-daq}
There are two primary sources of angle encoding inaccuracy: the timing jitter of the data acquisition system and noise in the encoder readout. The total noise should correspond to less than the 3 $\mu\mathrm{rad}\sqrt\mathrm{s}$ (Sec. \ref{req-daq}).

\begin{figure}
    \centering
    \includegraphics[width = 0.5\textwidth]{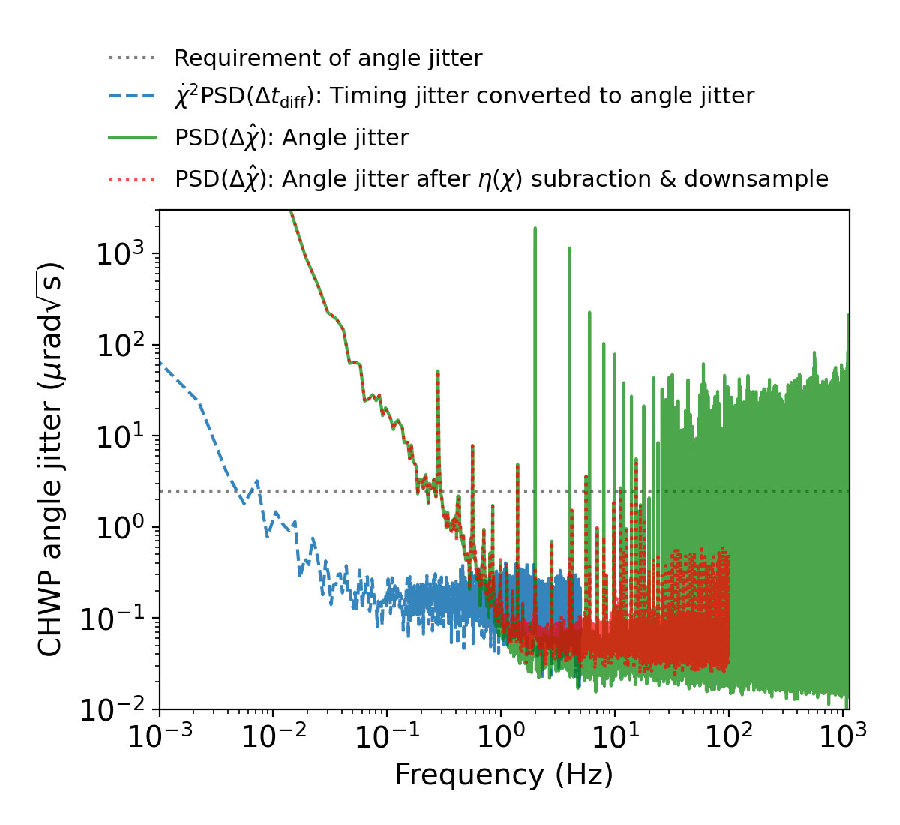}
    \caption{A measurement of the angle encoder performance over 1 hour. The CHWP is constantly rotating at 2 Hz. 
    The high frequency peaks in the raw PSD($\Delta\hat{\chi}$) are due to the non-uniformity of the encoder slot pattern, $\eta(\chi)$. The residual peaks after the $\eta(\chi)$ subtraction and down-sampling correspond to true fluctuation in the rotation angle, which are caused by vibrations of the PTC and the effects of the PID rotation control. The requirement applies only to the white noise level of the angle jitter.}
    \label{fig:angle_noise}
\end{figure}

We first discuss timing jitter, and more specifically the uncertainty of CHWP angle timestamp assignment.
A BeagleBone Black microcontroller\footnote{BeagleBone Black, Beagle Board: \protect\url{https://beagleboard.org/black}} acquires the encoder data, 
and assigns their timestamps based on its internal 200~MHz free-running clock (Sec. \ref{des-daq}).
Since the internal clock has long-timescale frequency drifts, the timestamps must be corrected by referencing the microcontroller to the SAT’s master clock, which also synchronizes with detector timestreams.
The microcontroller receives the master clock signal as IRIG-B\cite{IRIG} frame and position identifiers at a rate of 10 times per second. There are uncertainties, including jitter and data loss, in the measurement of the encoder and IRIG-B pulses, therefore we determine the CHWP angle after data acquisition by analyzing the full history of the internal timestamps for the two signals.
The corrected timestamp is expressed as
\begin{align}
    \hat{t} &= t_\mathrm{true} + \Delta t_\mathrm{clk},
    \label{eq:corrected_time}
\end{align}
where $\Delta t_\mathrm{clk}$ is the uncertainty in the correction after synchronizing the encoded angle timestamps to IRIG-B time. 
The effect of timing jitter on angle encoding accuracy is evaluated by the power spectral density (PSD) of $\Delta t_\mathrm{clk}$, which is determined by the relative time difference between the internal clock and IRIG-B time:
\begin{align}
    \label{eq:clk_diff}
    \Delta t_\mathrm{diff} = t_\mathrm{bbb}|_{t=t_\mathrm{true}-\Delta t_\mathrm{IRIG}} - t_\mathrm{true} 
    &\simeq \Delta t_\mathrm{bbb} - \Delta t_\mathrm{IRIG},
\end{align}
where $t_\mathrm{bbb}$ is the internal clock time, $\Delta t_\mathrm{bbb}$ is the drift of internal clock relative to IRIG-B, and $\Delta t_\mathrm{IRIG}$ is the timing jitter resulting from the detection of IRIG-B pulses.
Since $\Delta t_\mathrm{bbb}$ and $\Delta t_\mathrm{IRIG}$ are independent, the PSD is 
\begin{align}
    \label{eq:clk_jitter}
    \mathrm{PSD}(\Delta t_\mathrm{diff}) 
    &= \mathrm{PSD}(\Delta t_\mathrm{bbb}) +  \mathrm{PSD}(\Delta t_\mathrm{IRIG}) \nonumber \\
    &=  A f^{-\alpha} + \sigma_\mathrm{IRIG}^2/10~~~(\mathrm{sec}^2\cdot\mathrm{sec}),
\end{align}
where $A$ and $\alpha$ are the $1/f$ noise parameters of $\Delta t_\mathrm{bbb}$ and $\sigma_\mathrm{IRIG}^2$ is the noise variance in terms of the 10~Hz IRIG-pulse detection. 
The blue dashed curve in Fig.~\ref{fig:angle_noise} shows the corresponding angle jitter due to the timing jitter constructed from Eq.~\ref{eq:clk_jitter} multiplied by the angular velocity. The $1/f$ noise is given by $\Delta t_\mathrm{bbb}$ and the white noise level is determined by $\Delta t_\mathrm{IRIG}$. 
The corrected time, $\hat{t}$ (Eq.~\ref{eq:corrected_time}), is constructed by synchronizing the internal clock to the IRIG-B signal at a frequency of $f_\mathrm{sync}$ by linear interpolation.
Therefore, the PSD of the uncertainty in the corrected time $\Delta t_\mathrm{clk}$ is approximately expressed as
\begin{align}
    \mathrm{PSD}(\Delta t_\mathrm{clk}) &\simeq 
    \begin{cases}
    \sigma_\mathrm{IRIG}^2/10 & f \leq f_\mathrm{sync}/2 \\
    A f^{-\alpha} & f \geq f_\mathrm{sync}/2.
    \end{cases}
\end{align}
At frequencies below the synchronization Nyquist frequency ($f_\mathrm{sync}/2$), timestamp correction eliminates $1/f$ noise and the white noise of the IRIG detection jitter limits the timing accuracy. 
At frequencies above $f_\mathrm{sync}/2$ the timing accuracy still relies on the microcontroller's internal clock, resulting in the small amount of time drift.
As can be seen from Fig.~\ref{fig:angle_noise}, the angle jitter due to the timing jitter of the corrected time satisfies the requirement at all frequencies.

Next, we evaluate the encoder readout noise.
We place an upper limit on encoder readout noise using the encoder data and the CHWP's smooth rotation. 
The encoded CHWP angle is
\begin{align}
    \hat{\chi}(t_\mathrm{true}) &= \chi_\mathrm{true}(\hat{t}+\Delta t_\mathrm{enc}) + \eta(\chi_\mathrm{true}),
\end{align}
where $\hat{\chi}$ is the encoded rotation angle of the CWHP, $\chi_\mathrm{true}$ is the true rotation angle of the CHWP, and $\Delta t_\mathrm{enc}$ is the timing jitter that arises from the detection of photo-encoder pulses. 
$\eta(\chi_\mathrm{true})$ is the non-uniformity of the encoder slot pattern.
The CHWP angle jitter is defined by subtracting the smooth rotation:
\begin{align}
    \Delta\hat{\chi} &= \hat{\chi}(t_\mathrm{true}) -\overline{d\hat{\chi}/dt_\mathrm{true}}\cdot t_\mathrm{true}\nonumber \\
    &\simeq \Delta\chi_\mathrm{true} + 
    \dot{\chi}\Delta t_\mathrm{clk} + \dot{\chi}\Delta t_\mathrm{enc} + \eta(\chi_\mathrm{true}),
\end{align}
where $\Delta\chi_\mathrm{true}$ is the true drift of the CHWP angle that we would like to measure.
Since $\Delta\chi_\mathrm{true}$, $\Delta t_\mathrm{clk}$, $\Delta t_\mathrm{enc}$ and $\eta(\chi_\mathrm{true})$ are independent, the PSD of the CHWP angle jitter is 
\begin{align}
    \mathrm{PSD}(\Delta\hat{\chi})
    &\simeq \mathrm{PSD}(\Delta\chi_\mathrm{true})
    +\dot{\chi}^2\mathrm{PSD}(\Delta t_\mathrm{clk})
    \nonumber \\ &~~~~
    +\dot{\chi}^2\mathrm{PSD}(\Delta t_\mathrm{enc})
    +\mathrm{PSD}(\eta(\chi_\mathrm{true})),
    \label{eq:angle_jitter}
\end{align}
where
\begin{align}
    \label{eq:enc_jitter}
    \dot{\chi}^2\mathrm{PSD}(\Delta t_\mathrm{enc}) &\simeq 
    \frac{\dot{\chi}^2\sigma_\mathrm{enc}^2}{f_\mathrm{HWP}\times1140}~(\mathrm{rad}^2\cdot\mathrm{sec})
\end{align}
and $\sigma_\mathrm{enc}^2$ is the noise variance in terms of the encoder-pulse detection timing. There are 1140 pulses per revolution.
The green solid curve in Fig.~\ref{fig:angle_noise} shows Eq.~\ref{eq:angle_jitter} when the CHWP rotates at 2~Hz. 
The $1/f$ component is the true drift of the CHWP angle, $\Delta\chi_\mathrm{true}$. The white noise level is determined by $\Delta t_\mathrm{enc}$ and is well below the requirement.
The peaks at the harmonics of $f_\mathrm{HWP}$ arise from $\eta(\chi_\mathrm{true})$. 
$f_\mathrm{sync}$ is chosen to be 1~Hz to ensure that the timing jitter becomes subdominant to the drift of the CHWP angle at all frequencies.

To demodulate the TES detector timestream sampled at 200~Hz, the raw CHWP angles sampled at $1140\times f_\mathrm{HWP}$~Hz need to be down sampled.
In this procedure, the peaks at the higher harmonics of $f_\mathrm{HWP}$ must be subtracted so as not to contaminate the down-sampled angular timestream.
The non-uniformity of the encoder slot pattern, $\eta(\chi_\mathrm{true})$, is estimated by time averaging $\Delta\hat{\chi}$ for each angle step of the encoder signal as
\begin{align}
    \eta(\chi_\mathrm{true}) \simeq 
    \overline{ \Delta\hat{\chi}}~~\mathrm{for~each}~(\hat{\chi}~\mathrm{mod}~2\pi).
\end{align}
The red dotted curve in Fig.~\ref{fig:angle_noise} shows the angle jitter after the slot pattern subtraction and down-sampling.
The residual peaks at the harmonics of $f_\mathrm{HWP}$ are due to 
the slow modulation of the rotation-synchronous fluctuations of $\dot{\chi}$.
The achieved noise level is 0.07~$\mu\mathrm{rad}\sqrt\mathrm{s}$ , which is more than one order of magnitude lower than the requirement. 

\section{Conclusion}
\label{sec:conclusion}
We have presented the requirements, design and performance of the CHWP rotation mechanism for the Simons Observatory Small Aperture Telescopes.
This work advances the field of cryogenic polarization modulators for mm-wave and sub-mm astronomical observations by introducing the largest diameter CHWP constructed to date. 
The aperture size of the CHWP system was previously constrained by the size of the rotor's magnet ring. By overcoming the manufacturing limitations of the SMB and optimizing the optical design, the size of the system is now limited by the largest available size of sapphire plate.\footnote{Guizhou Haotian Optoelectronics Co., Ltd.} 
This work has also advanced the CHWP rotation drive techniques to improve operational efficiency, and introduced new methodologies for characterizing CHWP systems, which enabled improvement of the SMB to reduce vibration and evaluate the rotor displacement.

The CHWP for LF frequency is currently under development and will have different design parameters. The center alignment of the rotor will be more challenging since the LF HWP sapphire stack is a factor of $3\sim 4$ thicker, and thus $\sim 15$~kg heavier than the MF HWP. 
On the other hand, the LF band enjoys significantly less atmospheric fluctuations than the higher frequency bands, and thus the requirements on the CHWP rotation frequency can be relaxed.

Three CHWPs have been built and evaluated for three SATs, and three additional CHWPs are planned to be built.
CHWP performance satisfies all requirements, including a 478 mm clear aperture, a rotor temperature of <~70~K, a stator temperature of <~60~K, <~1.6~W of dissipation during continuous operation, rotation frequencies up to 3~Hz, rotation stability within 5~mHz, rotor alignment within 5~mm, and 0.07~$\mu\mathrm{rad}\sqrt\mathrm{s}$ of the encoded angle noise.
The presented CHWPs are expected to be deployed to the Chilean observation site and to see first light in 2023.
This development contributes not only to the SO project but also to the design and trade study for future experiments such as CMB-S4. \cite{abazajian_cmb-s4_2016, abitbol_cmb-s4_2017} 

\begin{acknowledgments}
The presented CHWP development was supported by JSPS KAKENHI Grant Numbers JP18H01240, JP19H00674, JP19K14732, JP21J11179, JP22H04913, JP23H00105, JP23H01202, 
and JSPS Core-to-Core program Grant Number JPJSCCA20200003,
and World Premier International Research Center Initiative (WPI), MEXT, Japan, and International Research Center Formation Program to Accelerate Okayama University Reform (RECTOR).
Work at LBNL is supported in part by the U.S. Department of Energy, Office of Science, Office of High Energy Physics, under contract No. DE-AC02-05CH11231,  
and a grant from the Simons Foundation under Award \#457687, B.K.,
and a grant from the Gordon and Betty Moore Foundation under Grant Number GBMF7939.
K.Y. acknowledges the support from XPS, WINGS Programs, the University of Tokyo. 
J.S. acknowledges the support from the International Graduate Program for Excellence in Earth-Space Science (IGPEES) and the JSR Fellowship, the University of Tokyo. 
D.S. acknowledges the support from FoPM, WINGS Program, the University of Tokyo. 
We thank Paul Barton at the Lawrence Berkeley National Laboratory for designing the motor drive electronics.
We thank Sean Adkins for his advice and assistance in improving the CHWP control system.
We thank the reviewers, whose suggestions clarify discussions throughout the paper.
\end{acknowledgments}

\tocless{\section*{Author Declarations}}

\tocless{}{\subsection*{Conflict of Interest}}
The authors have no conflicts to disclose.\\
\tocless{}{\subsection*{Author Contributions}}
\textbf{K.~Yamada}: Methodology (equal); Resources (equal); Investigation (equal); Validation (equal); Software (equal); Writing – original draft (lead).
\textbf{B.~Bixler}: Methodology (equal); Resources (equal); Investigation (equal); Validation (equal); Software (lead); Writing – original draft (supporting).
\textbf{Y.~Sakurai}: Conceptualization (equal); Methodology (equal); Resources (equal); Investigation (equal); Validation (equal); Software (equal); Funding Acquisition (supporting); Writing – original draft (equal).
\textbf{P.~C.~Ashton}: Conceptualization (equal); Methodology (equal); Resources (equal); Investigation (equal); Validation (equal); Writing – original draft (equal).
\textbf{J.~Sugiyama}: Methodology (equal); Resources (equal); Investigation (equal); Validation (equal); Software (supporting); Writing – original draft (supporting).
\textbf{K.~Arnold}: Project Administration (equal); Supervision (equal); Funding Acquisition (supporting).
\textbf{J.~Begin}: Investigation (supporting); Validation (supporting); Writing – review \& editing (supporting).
\textbf{L.~Corbett}: Investigation (supporting); Validation (supporting).
\textbf{S.~Day-Weiss}: Investigation (supporting); Validation (supporting); Writing – original draft (supporting).
\textbf{N.~Galitzki}: Supervision (supporting); Investigation (supporting); Validation (supporting); Writing – review \& editing (supporting).
\textbf{C.~A.~Hill}: Conceptualization (equal); Methodology (supporting); Software (equal); Writing – review \& editing (supporting).
\textbf{B.~R.~Johnson}: Supervision (supporting); Funding Acquisition (supporting); Writing – review \& editing (supporting).
\textbf{B.~Jost}: Writing – review \& editing (supporting).
\textbf{A.~Kusaka}: Project Administration (equal); Supervision (lead); Conceptualization (equal); Investigation (supporting); Validation (supporting); Funding Acquisition (lead); Writing – original draft (supporting).
\textbf{B.~J.~Koopman}: Software (equal).
\textbf{J.~Lashner}: Software (equal).
\textbf{A.~T.~Lee}: Project Administration (equal); Conceptualization (supporting); Supervision (supporting); Funding Acquisition (supporting).
\textbf{A.~Mangu}: Investigation (supporting); Validation (supporting).
\textbf{H.~Nishino}: Software (equal); Writing – review \& editing (supporting).
\textbf{L.~A.~Page}: Project Administration (equal); Conceptualization (supporting); Supervision (supporting); Investigation (supporting); Validation (supporting); Funding Acquisition (supporting); Writing – review \& editing (supporting).
\textbf{M.~J.~Randall}: Investigation (supporting); Validation (supporting).
\textbf{D.~Sasaki}: Investigation (supporting); Software (supporting); Writing – review \& editing (supporting). 
\textbf{X.~Song}: Investigation (supporting); Validation (supporting).
\textbf{J.~Spisak}: Investigation (supporting); Validation (supporting); Writing – original draft (supporting).
\textbf{T.~Tsan}: Investigation (supporting); Validation (supporting).
\textbf{Y.~Wang}: Investigation (supporting); Validation (supporting).
\textbf{P.~A.~Williams}: Investigation (supporting); Validation (supporting).
\\
\tocless{\section*{Data availability}}
The data that support the findings of this study are available from the corresponding author upon reasonable request.

\appendix
\section{Scan synchronous modulation of HWP rotation}
\label{sec:scan-mod}
\begin{figure}
    \centering
    \includegraphics[width = 0.48\textwidth]{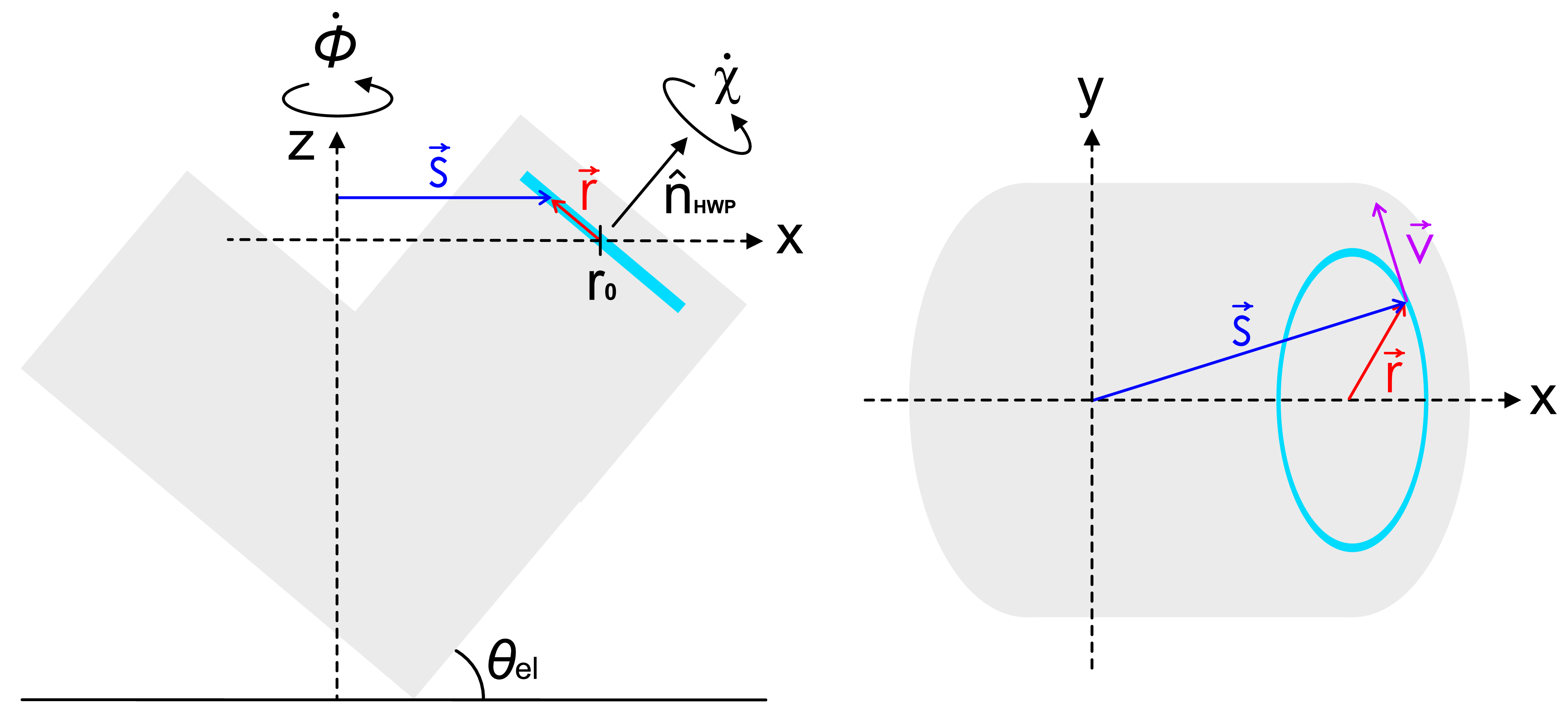}
    \caption{Schematic diagram of the rotating CHWP on the scanning telescope. The gray area represents a telescope's receiver, while the light blue circle represents the rotor.}
    \label{fig:scan_mod_sketch}
\end{figure}

Here we derive the scan modulation of the angular velocity of the CHWP, (Eq.~\ref{eq:scan_modulation}).
Figure \ref{fig:scan_mod_sketch} shows the schematic diagram of the rotating CHWP on the scanning telescope.
The coordinate system can be taken as shown, without the loss of generality.
We define $r$ and $\theta$ as the cylindrical coordinates of the rotor and $\rho(r, \theta)$ as the rotor mass density, which is symmetric about $\theta$.
The angular momentum of the rotating CHWP is
\begin{align}
    L_{\mathrm{HWP}} = \dot{\chi}\iint r~drd\theta~\rho r^2 = \dot{\chi} I,
\end{align}
where $I$ is the moment of inertia of the rotor.

Next, we calculate the angular momentum induced by the scanning telescope.
Let $r_0$ be the radial distance from the axis of the scan to the center of the rotor.
We assume that the CHWP bearing is infinitely rigid. 
This means that the net angular momentum contribution from the telescope scan is only along the direction of the rotor’s rotation axis. 
As shown in Fig. \ref{fig:scan_mod_sketch}, the angular momentum at each point of the rotor induced by the telescope scan is 
\begin{widetext}
\begin{align}
    \delta L_\mathrm{scan}
    &= \hat{n}_{\mathrm{HWP}}\cdot \left( \overrightarrow{r} \times \rho\overrightarrow{v} \right) \nonumber \\
    &= \begin{pmatrix} \cos\theta_\mathrm{el} \\ 0 \\ \sin\theta_\mathrm{el} \end{pmatrix} 
    \cdot \left\{ \begin{pmatrix} r\sin\theta_\mathrm{el}\cos\theta \\  r\sin\theta \\ -r\cos\theta_\mathrm{el} \end{pmatrix}
    \times \rho\dot{\phi}\begin{pmatrix} -r\sin\theta \\ r_0 + r\sin\theta_\mathrm{el}\cos\theta \\ 0 \end{pmatrix} \right\} \nonumber \\
    &= \rho\dot{\phi} \left( r_0r\cos\theta + r^2\sin\theta_\mathrm{el}\right),
\end{align}
\end{widetext}
where $\phi$ and $\theta_\mathrm{el}$ are the azimuth and elevation angles of the telescope, respectively.
The total angular momentum is obtained by integrating over the rotor as
\begin{align}
    L_\mathrm{scan} 
    = \iint r~drd\theta~\delta L_\mathrm{scan}
    = \dot{\phi}\sin\theta_\mathrm{el} I.
\end{align}
From the conservation law of angular momentum, the following equation holds.
\begin{align}
    L_\mathrm{scan} + L_\mathrm{HWP} = \mathrm{const.}
\end{align}
Thus, Eq. \ref{eq:scan_modulation} holds.


\section{Measurement of the displacement of rotor} \label{sec:sag-method}
\begin{figure}
    \centering
    \includegraphics[width = 0.48\textwidth]{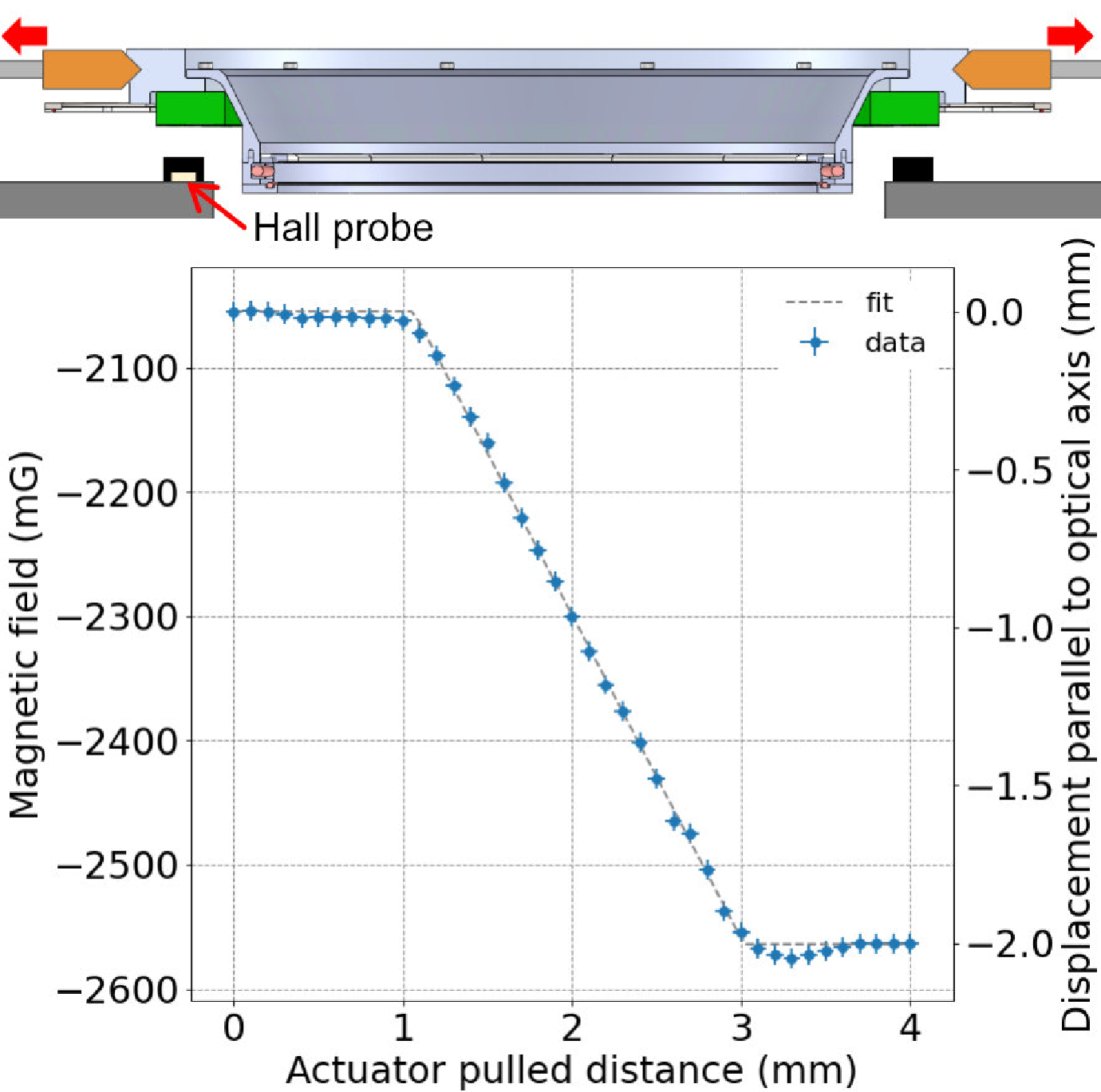}
    \caption{Top panel: schematic diagram of the measurement setup for the rotor displacement along the optical axis. The Hall probe is located right below the magnet ring to measure the relative displacement of the rotor. Grippers are completely clear of the rotor when retracted 9~mm. 
    Bottom panel: measurement of the displacement along the optical axis.
    From 0-1~mm, the rotor is not moving due to the play in the joints of the grippers.
    From 1-3~mm, the rotor is sliding down the wedge of the gripper heads.
    From 3~mm onwards, the rotor is floating.}
    \label{fig:z-sag}
\end{figure}

Here we elaborate on the methods used to make the displacement measurements summarized in Sec.~\ref{perf-sag}.
Displacement measurements along the optical axis are made using the gripping mechanisms (Sec.~\ref{des-gripper}) and a Hall probe (Sec.~\ref{des-daq}) installed below the rotor’s magnet ring. The measurement begins by pointing the receiver at an elevation of 90$^\circ$ and fully gripping the rotor. The actuators controlling the gripper positions are then retracted, allowing the rotor to sink along the optical axis. The gripper head geometry is designed such that retraction distance corresponds directly to the rotor displacement, and changes in the Hall probe’s measured field are monitored. The total displacement of the rotor is estimated from the distance the grippers have moved while the Hall probe changes linearly.
Figure~\ref{fig:z-sag} shows a schematic of the measurement setup and the measured magnetic field of the rotor magnet as the grippers are gradually retracted. 
The systematic error of the displacement is $\pm$0.3~mm, resulting from changes in the Hall probe sensitivity and the rotor magnet’s field caused by temperature drift during the measurement.

\begin{figure}
    \centering
    \includegraphics[width = 0.48\textwidth]{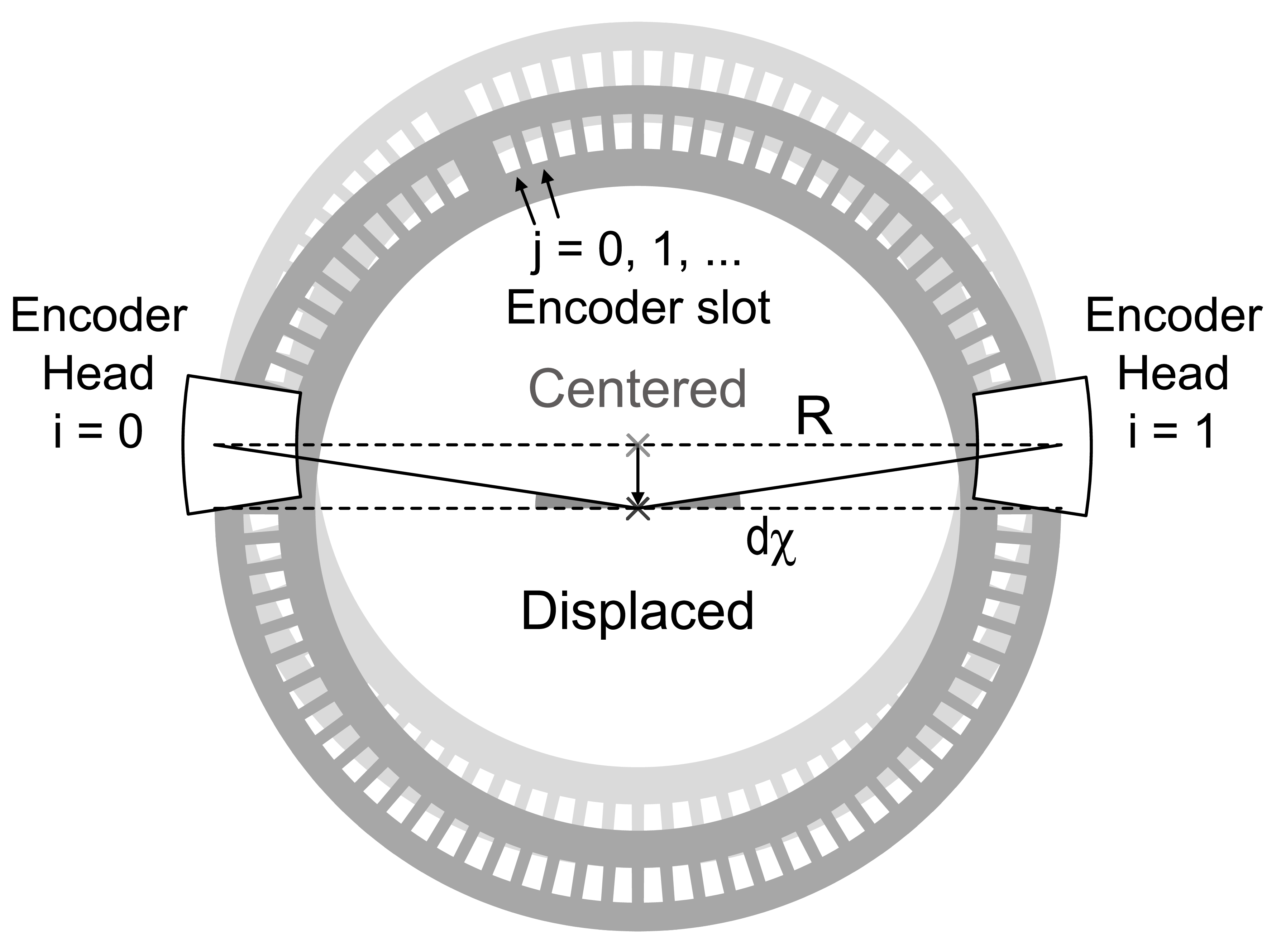}
    \caption{The schematic diagram of the measurement setup of the off-center displacement. When the rotor is off-centered perpendicular to the line connecting the encoders, the pair incorrectly measures the rotation angle by $d\chi$. The off-center displacement can be calculated by Eq.~\ref{eq:sag_vs_angle}. }
    \label{fig:sag}
\end{figure}

The off-center displacement is calculated by the measured additional angle offset between a pair of encoders placed 180$^\circ$ from one another as shown in Fig.~\ref{fig:sag}. The additional angle offset is
\begin{align}
    \label{eq:phasediff}
    d\chi_j &= \pi f_\mathrm{HWP\ j} \left(t^0_j-t^1_{j+570}\right),
\end{align}
where $t^i_j$ is the timing when i-th encoder reads the j-th encoder slot and
\begin{align}
    f_\mathrm{HWP\ j} &= \frac{1}{1140(t^{0}_j-t^{0}_{j+1})}= \frac{1}{1140(t^{1}_j-t^{1}_{j+1})}
\end{align}
is the rotation frequency.
Since the total number of encoder slots is 1140, the slot 570 positions away is at the opposite side of the encoder plate. If the CHWP is completely centered, $d\chi_j$ is always zero. If there is an off-center displacement of the rotor perpendicular to the line connecting the encoders, 
diametrically opposite slots will travel different distances between detection, resulting in a time delay that can be converted to the angle offset, $d\chi_j$.
The pair of encoders are synchronized by the shared IRIG-B reference signal described in Sec.~\ref{perf-daq}. 
The displacement perpendicular to the line connecting the pair of encoders is thus
\begin{align}
    \label{eq:sag_vs_angle}
    \mathrm{Displacement}_j &= R\times\tan(d\chi_j),
\end{align}
where $R = 334$~mm is the radius of the encoder slots.

The measurement accuracy of the encoder angle shift due to off-center displacement is 0.02$^\circ$, which is negligible compared to the required accuracy of the polarization angle calibration. The SAT platform will rotate the boresight $\pm$ 60$^\circ$. Since the phase shift between the pair of encoders is at its maximum when the boresight angle is 0$^\circ$, the requirement of the accuracy of the polarization angle is satisfied at any boresight angle.

\section{Precautions}
\label{sec:failure-modes}
Successful operation of the CHWP requires paying careful attention to a number of system characteristics. This section presents a summary of the precautions we learned to take while building the CHWP and evaluating its performance.

Physical interference of the rotor is the most common inhibitor of CHWP rotation. 
This interference can be caused by misplaced or loose screws, nuts, tape, motor sprocket magnets, encoder or motor cables, or multi-layer insulation.
To avoid the risk of physical interference, use of nuts and tape are minimized, screws are secured with thread-locker (LOCTITE 263) and the motor sprocket magnets are secured with epoxy (Stycast 2850 FT).
Moreover, the magnetic screws or mechanical parts, including 304 stainless steel, must not be used. This is because the rotor's magnet ring is strong enough to dislodge them and cause physical interference. 

Care must also be taken in the operation of electrical devices, including the encoders, motors, and grippers. Protective measures for electrical devices are crucial, such as the protection circuit for the optical encoders, the current limit function for the motor drive power source, and the limit switch for the grippers.
Finally, drastic temperature changes must be avoided while the encoder LEDs are biased, as exposure to rapidly changing environments can cause degradation.


\begin{thebibliography}{98}
\bibitem [{Notexv()}]{Notexv}%
  \BibitemOpen
  \bibinfo {note} {The minimum elevation angle of the SAT platform is 20$^\circ
  $.}\BibitemShut {Stop}%
\end{thebibliography}
\end{document}